\documentclass[12pt,usenames,dvipsnames]{article}

\usepackage{latexsym}
\usepackage{amssymb,amsfonts,amsmath}
\usepackage{graphicx} 
\usepackage{indentfirst}
\usepackage{bbm}
\usepackage{amssymb}
\usepackage{verbatim}
\usepackage{amsmath, amsthm,amssymb}
\usepackage{mathrsfs}
\usepackage{hyperref}
\usepackage{amsfonts}
\usepackage{dsfont}
\usepackage{cite}
\usepackage{xcolor}
\usepackage{enumerate}
\usepackage{cleveref}
\usepackage{cjhebrew}
\usepackage{arabtex}

\parskip=\medskipamount

\newcommand {\cD}{{\cal D}}
\newcommand {\cE}{{\cal E}}

\newcommand {\cG}{{\cal G}}
\newcommand {\cH}{{\cal H}}

\newcommand {\cK}{{\cal K}}
\newcommand {\cL}{{\cal L}}
\newcommand {\cM}{{\cal M}}
\newcommand {\cN}{{\cal N}}
\newcommand {\cO}{{\cal O}}

\newcommand {\cR}{{\cal R}}
\newcommand {\cS}{{\cal S}}
\newcommand {\cT}{{\cal T}}

\newcommand {\cX}{{\cal X}}

\def\a{\alpha}

\def\b{\beta}

\def\d{\delta}
\def\e{\epsilon}
\def\f{\phi}
\def\g{\gamma}
\def\G{\Gamma}
\def\i{\iota}

\def\k{\kappa}
\def\l{\lambda}
\def\m{\mu}
\def\n{\nu}
\def\o{\omega}
\def\p{\pi}
\def\q{\theta}
\def\r{\rho}
\def\s{\sigma}
\def\t{\tau}

\def\x{\xi}
\def\z{\zeta}

\def\F{\Phi}

\def\L{\Lambda}
\def\O{\Omega}
\def\P{\Pi}

\def\S{\Sigma}
\def\U{\Upsilon}
\def\X{\Xi}

\def\rd{{\rm d}}
\def\ri{{\rm i}}
\def\re{{\rm e}}

\def\N{{\cal N}}

\newcommand{\ah}{{\hat{\a}}}
\newcommand{\bh}{{\hat{\b}}}
\newcommand{\gh}{{\hat{\g}}}
\newcommand{\dhat}{{\hat{\d}}}
\newcommand{\mh}{{\hat{\m}}}

\newcommand{\ad}{{\dot{\alpha}}}                           
\newcommand{\bd}{{\dot{\beta}}}                            
\newcommand{\ve}{\varepsilon}                            

\newcommand{\ab}{{\a\b}}

\newcommand{\vf}{\varphi}

\newcommand{\be}{\begin{equation}}
\newcommand{\ee}{\end{equation}}
\newcommand{\bea}{\begin{eqnarray}}
\newcommand{\eea}{\end{eqnarray}}

\def\dt#1{{\buildrel {\hbox{\LARGE .}} \over {#1}}}    

\newcommand{\bm}[1]{\mbox{\boldmath$#1$}}

\def\double #1{#1{\hbox{\kern-2pt $#1$}}}

\newcommand{\hm}{{\hat{m}}}

\newcommand{\ha}{{\hat{a}}}
\newcommand{\hb}{{\hat{b}}}
\newcommand{\hc}{{\hat{c}}}
\newcommand{\hd}{{\hat{d}}}

\newcommand{\hM}{{\hat{M}}}

\newcommand{\hA}{{\hat{A}}}
\newcommand{\hB}{{\hat{B}}}
\newcommand{\hC}{{\hat{C}}}

\newcommand{\hal}{{\hat{\a}}}

\newcommand{\gd}{{\dot\g}}

\newif\ifdtup

\newcommand{\bsubeq}{\begin{subequations}}
\newcommand{\esubeq}{\end{subequations}}

\numberwithin{equation}{section}

\newcommand{\sSU}{\mathsf{SU}}
\newcommand{\sSL}{\mathsf{SL}}
\newcommand{\sGL}{\mathsf{GL}}
\newcommand{\sSO}{\mathsf{SO}}
\newcommand{\sO}{\mathsf{O}}
\newcommand{\sU}{\mathsf{U}}

\newcommand{\sSpin}{\mathsf{Spin}}

\newcommand{\sUSp}{\mathsf{USp}}

\newcommand{\id}{\mathds{1}}

\newcommand{\ua}{{\underline{a}}}
\newcommand{\ub}{{\underline{b}}}

\newcommand{\ui}{{\underline{i}}}
\newcommand{\uj}{{\underline{j}}}
\newcommand{\uk}{{\underline{k}}}
\newcommand{\ul}{{\underline{l}}}

\newcommand{\T}{\text{T}}

\newcommand{\braket}[2]{\langle #1 | #2 \rangle}

\newcommand{\tb}[1]{\textbf{#1}}
\newcommand{\tmu}{{\tilde{\m}}}
\newcommand{\tnu}{{\tilde{\n}}}
\newcommand{\tro}{{\tilde{\r}}}
\newcommand{\tsi}{{\tilde{\s}}}
\newcommand{\ti}{{\tilde{i}}}
\newcommand{\tj}{{\tilde{j}}}

\begin{document}
	
\begin{titlepage}
	\begin{flushright}
	June, 2024 \\
	Revised version: February, 2025
	\end{flushright}
	\vspace{5mm}
	
	\begin{center}
		{\Large \bf 
			Embedding formalism for AdS superspaces in five dimensions
		}
	\end{center}

\begin{center}
		{\bf Nowar E. Koning and Sergei M. Kuzenko} \\
		\vspace{5mm}
		
		\footnotesize{ 
			{\it Department of Physics M013, The University of Western Australia\\
				35 Stirling Highway, Perth W.A. 6009, Australia}}  
		~\\
		\vspace{2mm}
		~\\
		Email: \texttt{nowar.koning@research.uwa.edu.au,
			sergei.kuzenko@uwa.edu.au }\\
		\vspace{2mm}
	\end{center}
	
\begin{abstract}
		\baselineskip=14pt
The standard geometric description of $d$-dimensional anti-de Sitter (AdS) space is 
a quadric in ${\mathbb R}^{d-1,2}$ defined by $(X^0)^2 - (X^1)^2 - \dots - (X^{d-1})^2 + (X^d)^2 = \ell^2 = \text{const}$. In this paper we provide a supersymmetric generalisation of this embedding construction in the $d=5$ case. Specifically, a bi-supertwistor realisation is given for the ${\cal N}$-extended AdS superspace $\text{AdS}^{5|8\cal N}$, with ${\cal N}\geq 1$. The proposed formalism offers a simple construction of AdS super-invariants. As an example, we present a new model for a massive superparticle in $\text{AdS}^{5|8\cal N}$ which is manifestly invariant under the AdS isometry supergroup $\mathsf{SU}(2,2|{\cal N})$ and involves two independent two-derivative terms.
\end{abstract}
\vspace{5mm}
	
	\vfill
	
	\vfill
\end{titlepage}

\newpage
\renewcommand{\thefootnote}{\arabic{footnote}}
\setcounter{footnote}{0}

\tableofcontents{}
\vspace{1cm}
\bigskip\hrule

\allowdisplaybreaks

\section{Introduction}

As is well-known, the algebraic underpinning for the AdS/CFT duality is the dual role played by the group $\sSO_0(d-1,2)$. On the one hand, it is the connected conformal group of the $(d-1)$-dimensional compactified Minkowski space $\overline{\mathbb M}^{d-1}= (S^{d-2} \times S^1)/{\mathbb Z}_2$.\footnote{Strictly speaking,  the connected conformal group should be identified with $\sSO_0 (d-1, 2)/{\mathbb Z}_2$ 
 if $d$ is odd.} On the other hand, it is the connected component of the isometry group of AdS$_d$, the $d$-dimensional AdS space. The situation in the supersymmetric world is analogous modulo the fact that superconformal groups exist only  in six or less space-time dimensions \cite{Nahm:1977tg}. In particular, for $d=5$ 
the supergroup $\sSU(2,2|\cN)$ is realised either 
(i) as the superconformal group in four dimensions; or (ii) as the $\cN$-extended AdS supergroup in five dimensions. 

From the group-theoretic point of view, an $\cN$-extended AdS superspace in five dimensions is defined as 
\begin{align} \label{n extended ads}
	\text{AdS}^{5|8\N} = \frac{\sSU(2,2|\N)}{\sSpin(4,1) \times \sU(\N)}\,,
\end{align}
see, e.g., \cite{Bandos:2002nn}. In this family of coset spaces, the only conformally flat superspace is 
\begin{align} \label{n=1 ads}
	\text{AdS}^{5|8} = \frac{\sSU(2,2|1)}{\sSpin(4,1) \times \sU(1)}~,
\end{align}
as was argued in \cite{Bandos:2002nn}.\footnote{No explicit construction of a conformally flat $\cN=1$ AdS superspace geometry was given in \cite{Bandos:2002nn}. Such a construction was derived in \cite{Kuzenko:2008kw} by making use of the superspace formulation for the 5D $\cN=1$ Weyl supermultiplet developed in \cite{Kuzenko:2008wr}.}
The $\cN=1$ case is exceptional since there is a unique superconformal algebra in five dimensions
\cite{Nahm:1977tg}, the so-called  $F(4)$ superalgebra discovered by Kac \cite{Kac}, corresponding to $\cN=1$. This implies that there is only one type of $d=5$ superconformal tensor calculus  which was
developed independently  by two groups: 
Fujita, Kugo, and Ohashi
\cite{Ohashi1,Ohashi2,Ohashi3,Ohashi4} 
and Bergshoeff et al. \cite{Bergshoeff1,Bergshoeff2,Bergshoeff3}. Finally, there is only one type of $d=5$ conformal superspace \cite{Butter:2014xxa}.

The simple AdS superspace, eq. \eqref{n=1 ads}, is a maximally symmetric background of the minimal $\cN=1$ supergravity geometry sketched by Howe\cite{Howe5Dsugra} (see also \cite{HL}) and fully developed in \cite{Kuzenko:2007cj, Kuzenko:2007hu}. 
It was shown in  \cite{Butter:2014xxa, Kuzenko:2014eqa}
that $\text{AdS}^{5|8} $ is the only maximally supersymmetric solution in the off-shell gauged $\cN=1$ supergravity with the following compensators: 
(i) the vector multiplet; and (ii) the $\cO(2) $ multiplet.
Supersymmetric nonlinear $\s$-models in AdS$_5$  have interesting off-shell \cite{Kuzenko:2007aj} and geometric properties \cite{Bagger:2011na}.

In the non-supersymmetric case, there exist two different but equivalent realisations of AdS$_d$: (i) as the coset space $\sO(d-1\,,2) / \sO(d-1\,,1)$; and (ii) as a hypersurface in $\mathbb{R}^{d-1\,,2}$
\begin{align} \label{ads embedding}
	-(X^{0})^{2} + (X^{1})^{2} + \cdots + (X^{d-1})^{2} - (X^{d})^{2} = -\ell^{2} = \text{const}\,.
\end{align}
Both realisations have found numerous applications in the literature. 
Supersymmetric analogues of \eqref{ads embedding} have recently been developed in three and four dimensions, see \cite{Kuzenko:2021vmh,Kuzenko:2023yak, Koning:2023ruq}. 
For the $d=5$ case, a bi-twistor realisation for (complexified) AdS$_5$ also exists, see, e.g., \cite{Sinkovics:2004fm, Adamo:2016rtr}.
To the best of our knowledge, a supersymmetric extension of \eqref{ads embedding} in five dimensions is yet to be developed in the literature, although a similar construction exists in the context of superstrings on AdS$_5 \times S^{5}$ \cite{Roiban:2000yy}.  
In this paper we develop the superembedding formalism\footnote{For a pedagogical review of superembeddings see \cite{Sorokin:1999jx, Bandos:2023web}.} for AdS$^{5|8\N}$. 

Since the work by Ferber \cite{Ferber:1977qx}, supertwistors have found numerous applications in theoretical and mathematical physics. 
In particular, supertwistor realisations of compactified $\N$-extended Minkowski superspaces have been developed in four \cite{Manin:1988ds, Kotrla:1984ky} and three \cite{Howe:1994ms, Kuzenko:2010rp} dimensions, and their harmonic/projective extensions have been derived \cite{Howe:1994ms, Kuzenko:2010rp, Rosly:1985nyf, Lukierski:1988vw, Hartwell:1994rp, Howe:1995md, Kuzenko:2006mv, K-compactified12, Buchbinder:2015qsa}.\footnote{Recently, there has been an interesting study of the geometric nature of the points at infinity of Minkowski superspace \cite{Boulanger:2023gpw}.}
Supertwistor realisations of AdS$^{(3|p,q)}$ and AdS$^{4|4\N}$ were provided in \cite{Kuzenko:2021vmh}. 
Recently, supertwistor formulations for conformal supergravity theories in diverse dimensions have been proposed \cite{Howe:2020xrg, Howe:2020hxi}. 
Furthermore, (super)twistor descriptions of (super)particles in AdS spaces have been studied extensively in the literature \cite{Claus:1999xr, Claus:1999zh, Claus:1999jj, Bandos:1999pq, Zunger:2000wy,Cederwall:2000km, Cederwall:2004cf, Arvanitakis:2016vnp, Arvanitakis:2017cpk, Uvarov:2018ose, Uvarov:2019vmd}.

This paper is organised as follows.
In section \ref{boson ads} we begin by introducing the twistor realisation for AdS$_5$.
Following that, we recall the standard bi-twistor realisation and prove their equivalence. 
In section \ref{susy ads} we describe a supertwistor realisation for the $\N$-extended AdS superspace in five dimensions, AdS$^{5|8\N}$. 
We then introduce a bi-supertwistor realisation for the same superspace and prove their equivalence. 
Finally we discuss an invariant two-point function on AdS$^{5|8\N}$. 
In section \ref{coset rep section} we develop coset representatives for both the bosonic and supersymmetric cases, and discuss their corresponding coordinate systems. 
Section \ref{geometry} is devoted to the study of the geometry of AdS$^{5|8\N}$.
Section \ref{projective} introduces a family of homogeneous superspaces of the form AdS$^{5|8\N} \times \mathbb{X}_{m}^{\N}$, where the internal space $\mathbb{X}_{m}^{\N}$ is realised in terms of $m \leq \N$ odd supertwistors. The main body of the paper is accompanied by appendix \ref{spin} which outlines our spinor conventions in diverse dimensions.


\section{AdS$_5$} \label{boson ads}

As is well-known, compactified Minkowski space $\overline{\mathbb M}{}^4 = (S^{3} \times S^1)/{\mathbb Z}_2$ can be realised \cite{Penrose1,Penrose2,Segal,Todorov,WW} 
as the set of null planes
in the twistor space $\mathfrak{T} $ which is a four-dimensional complex vector space, $\mathfrak{T} = {\mathbb C}^4$, endowed with the linear action of the conformal group $\sSU(2,2)$, 
\bea
\sSU(2,2) = \left\{ g \in \sSL(4,\mathbb{C})\,, \quad g^{\dag}\O g = \O \, ,\quad 
\O =  \left(\begin{array}{cc}
	0 & \id_{2} \\
	\id_{2} & 0
\end{array}\right)
\right\}~.
\label{su(2,2) master}
\eea
Twistors can also be used to describe AdS$_5$.

\subsection{Algebraic background}
A twistor $T \in {\mathfrak T}$ is a column vector
\begin{align}
	T = (T_{\ah}) = \left(\begin{array}{c}
		f_{\a} \\
		\bar{g}^{\ad}
	\end{array}\right)\,, 
	\qquad \a = 1\,,2\,, \quad \ad= \dot{1} \,, \dot{2}\,.
\end{align}
The group $\sSU(2,2)$  naturally acts on $\mathfrak T$ by transformations 
\begin{align}
	T_{\ah} \longrightarrow g_{\ah}{}^{\bh}T_{\bh}\,, \qquad g &= \big(g_{\ah}{}^{\bh}\big) \in \sSU(2,2)~.
\end{align}
A unique $\sSU(2,2)$ invariant inner product on $\mathfrak T$ is given by 
\begin{align}
	\braket{T}{S} := T^{\dag}\O S \equiv \bar{T} S\, ,
\end{align}
where we have defined
\begin{align}
	\bar{T}= T^\dagger \O = (\bar{T}^{\ah})\,, \qquad \bar{T}^{\ah} = (T_{\bh})^* \,\O^{\bh\ah}~, \qquad 
	\O &= \big(\O^{\ah\bh}\big) \, .
\end{align}
We will refer to $\bar T$ as the dual of $T$. It transforms as 
\begin{align}
	\bar{T}^{\ah} \longrightarrow \bar{T}^{\bh}(g^{-1})_{\bh}{}^{\ah}\,. 
\end{align}
The space of dual twistors will be denoted $\bar {\mathfrak T}$. 

In what follows, of special importance for our analysis will be 
a subgroup of $\sSU(2,2)$, denoted $\sUSp(2,2)$, consisting of those group elements which satisfy the additional condition 
\begin{align} \label{usp cond}
	g ^{\T} C g = C\,, \qquad C = (\ve^{\ah \bh} ) = \left(\begin{array}{cc}
		~\ve^{\a\b} ~ & 0 
		\\
		0 & -\ve_{\ad \bd}
	\end{array}\right)\,,
\end{align}
with $ \ve^{\ab}$ and $\ve_{\ad \bd}$ being defined in \eqref{epsilon def}. 
The Lie algebras of $\sSU(2,2)$ and $\sUSp(2,2)$ consist of elements of the form \eqref{su(2,2) generic algebra element} and \eqref{so(4,1) generic algebra element}, respectively, and are denoted $\mathfrak{su}(2,2)$ and $\mathfrak{usp}(2,2)$.


\subsection{Twistor realisation}

Now we turn to describing
the twistor realisation for AdS$_5$. 
Let us introduce a set of four linearly independent twistors 
\begin{align}
	T^\tmu = \big( T_{\ah}{}^{\tmu} \big)\,, \qquad \tmu = 1\,,2\,,3\,,4\,.
\end{align}
We further restrict our attention to those quartets $T^\tmu$  that satisfy the constraints 
\bsubeq \label{group cond with I}
\begin{align} 
	(T_{\ah}{}^\tmu )^* \O^{\ah\bh} T_{\bh}{}^{\tnu} &= \O^{\tmu\tnu}\,, \qquad (\O^{\tmu\tnu}) := \left(\begin{array}{cc}
		0 & \id_{2} \\
		\id_{2} & 0
	\end{array}\right)\,,
\\
 \det \left(T_{\ah}{}^{\tmu}\right) &= 1\,. \label{det const}
\end{align}
\esubeq
The space of all such quartets will be denoted $\mathfrak{F}$.

On the space $\mathfrak{F}$, we introduce the equivalence relation
\begin{align} \label{usp equiv}
	T_{\ah}{}^{\tmu} &\sim T_{\ah}{}^{\tnu} \l_{\tnu}{}^{\tmu}\,, \qquad \l \in \sUSp(2,2)\,.
\end{align}
We emphasise that the constraints \eqref{group cond with I} are invariant under the equivalence transformations \eqref{usp equiv}. 
The group $\sSU(2,2)$ acts on the space $\mathfrak{F}$ as 
\begin{align}
	T_{\ah}{}^{\tmu} \rightarrow g_{\ah}{}^{\bh}T_{\bh}{}^{\tmu}\,, \qquad g \in \sSU(2,2)\,. 
\end{align}
This action is naturally extended to the quotient space $\mathfrak{F} /\sim$. 
The latter proves to be a homogeneous space for the group $\sSU(2,2)$. 
It turns out that 
\begin{align} \label{ads twistor def}
	\text{AdS}_{5} = \mathfrak{F}/\sim\,. 
\end{align}

To prove \eqref{ads twistor def}, we introduce the bi-twistor 
\begin{align} \label{ads bitwistor}
	X_{\ah\bh} := \ell \mathfrak{C}_{\tmu\tnu}T_{\ah}{}^{\tmu}T_{\bh}{}^{\tnu} = -X_{\bh\ah}\,, 
\end{align}
where the matrix $\mathfrak{C}$ is defined as 
\begin{align}
	\mathfrak{C} = (\mathfrak{C}^{\tmu\tnu}) = \left(\begin{array}{cc}
		\ve & ~0~ 
		\\
		~0~ & -\ve^{-1}
	\end{array}\right)
	\,, \qquad
	\mathfrak{C}^{-1} = 
	(\mathfrak{C}_{\tmu\tnu}) 
	\,, \qquad
\ve = \left(\begin{array}{cc}
		0 & 1
		\\
		-1 & ~0~
	\end{array}\right)\,.
\end{align}
We present the matrix $\mathfrak{C}$ in this form for later convenience. 
The bi-twistor $X_{\ah\bh}$ is invariant under the equivalence relation \eqref{usp equiv}, and as such can be used to parametrise the space \eqref{ads twistor def}.

As shown in appendix \ref{spinor conventions}, a one-to-one correspondence 
exists between complex vectors $V^{\ua}$ in $4+2$ dimensions and antisymmetric bi-twistors $V_{\ah\bh} = -V_{\bh\ah}$. It is defined by the rule \eqref{o-to-o}, 
\begin{align}
	V^{\ua} ~\to~ V_{\ah\bh}  = V^{\ua}(\S_{\ua})_{\ah\bh}\,,
\end{align}
where the matrices $\S_{\ua}$ are given in \eqref{bt basis def}.
The six-vector corresponding to \eqref{ads bitwistor}, 
\begin{align}
	X^{\ua} = \frac{1}{4}(\tilde{\S}^{\ua})^{\ah\bh}X_{\ah\bh}\,,
\end{align}
is real, because the bi-twistor $X_{\ah\bh}$ satisfies the reality condition, 
\begin{align} \label{ads real bt}
	\bar{X}^{\ah\bh} := \O^{\ah\gh}(X^{\dag})_{\gh\dhat}\O^{\dhat\bh} = \frac{1}{2}\ve^{\ah\bh\gh\dhat}X_{\gh\dhat}\,,
\end{align}
see \eqref{index reality}.
In order to prove \eqref{ads real bt} we use the identities 
\bsubeq
\begin{align}
	T_{\ah}{}^{\tmu}T_{\bh}{}^{\tnu}T_{\gh}{}^{\tro}T_{\dhat}{}^{\tsi}\ve^{\ah\bh\gh\dhat} &= \ve^{\tmu\tnu\tro\tsi}\,,
	\\
	\mathfrak{C}^{\tmu\tnu}\mathfrak{C}^{\tro\tsi} + \mathfrak{C}^{\tmu \tro}\mathfrak{C}^{\tsi \tnu} + \mathfrak{C}^{\tmu \tsi}\mathfrak{C}^{\tnu \tro} &= - \ve^{\tmu\tnu\tro\tsi}\,,
\end{align}
\esubeq
the first of which is equivalent to $\det (T_{\ah}{}^{\tmu}) = 1$.
Making use of the completeness relation
\begin{align}
	(\tilde{\S}^{\ha})^{\ah\bh}(\tilde{\S}_{\ha})^{\gh\dhat} &= 2\ve^{\ah\bh\gh\dhat}\,,
\end{align}
see \eqref{completeness relations}, we obtain 
\begin{align}
	X^{\ua}X_{\ua} = -\ell^{2}\,. 
\end{align}


\subsection{Bi-twistor realisation} \label{bt realisation}

A bi-twistor realisation for (complexified) AdS$_5$ exists in the literature, see \cite{Sinkovics:2004fm, Adamo:2016rtr}.
In this section we will prove its equivalence with the twistor realisation described above. 

Let  $\mathcal{L}$ be the set of bi-twistors, $X_{\ah\bh} = -X_{\bh\ah}$, which obey the following constraints:
\bsubeq \label{bt properties}
\begin{align} 
	\bar{X}^{\ah\bh} &= \O^{\ah\gh}(X^{\dag})_{\gh\dhat}\O^{\dhat\bh} = \frac{1}{2}\ve^{\ah\bh\gh\dhat}X_{\gh\dhat}\,, \label{bt tilde bar con}
	\\
	\bar{X}^{\ah\gh}X_{\gh\bh} &= \ell^{2} \d^{\ah}{}_{\bh}\,, \label{delta con}
	\\
	X_{[\ah\bh}X_{\gh\dhat]} &= -\frac{1}{3} \ell^{2}\ve_{\ah\bh\gh\dhat}\
	\quad \Longleftrightarrow \quad \ve^{\ah\bh\gh\dhat}X_{\ah\bh}X_{\gh\dhat} = -8\ell^{2} \, .
	 \label{XX epsilon}
\end{align}
\esubeq
Our aim is to show that the space $\mathcal{L}$ can be identified with $\mathfrak{F} / \sim$. 
First of all, it is not difficult to see that $\mathcal L$ provides an equivalent realisation of AdS$_5$.
Making use of the correspondence \eqref{o-to-o}, we associate a unique six-vector $X^{\ua}$ with each $X_{\ah\bh} \in \cL\,.$
The condition \eqref{bt tilde bar con} means that the six-vector $X^{\ua}$ is real, whereas \eqref{delta con} recreates the AdS condition $X^{2} = -\ell^{2}$. 
We therefore can identify AdS$_5$ with $\cL$. 
It remains to show that every $\hat{X} = (X_{\ah\bh} ) \in \cL $ can be written in the form \eqref{ads bitwistor}, 
where the twistors $T^\tmu$ are constrained as in \eqref{group cond with I}.

An $\sSU(2,2)$ transformation acts on $\hat{X} = (X_{\ah\bh} ) $ as 
\begin{align}
	\hat{X} &\longrightarrow g \hat{X} g^{\T} = X'^{\ua} (\S_{\ua})\,, \qquad g \in \sSU(2,2)\,,
\end{align}
with $X'^{\ua}$ given by 
\begin{align}
	X'^{\ua} = \L^{\ua}{}_{\ub} X^{\ub}\,, \qquad \L \in \sSO_{0}(4,2)\,.
\end{align}
This transformation preserves the conditions \eqref{bt properties}. 
Now we make use of the fact that AdS$_5$ is a homogeneous space of $\sSO_{0}(4,2)$, and has a global coset representative $s(X)$, see, e.g., \cite{Kuzenko:1995aq}. That is, for any $X^{\ua} \in$ AdS$_5$ it holds that 
\begin{align}
	X^{\ua} = s(X)^{\ua}{}_{\ub}X_{(0)}^{\ub}\,, \qquad s(X) \in \sSO_{0}(4,2)\,,
\end{align}
for some base point $X_{(0)}^{\ua}$. 
We can choose the base point to be 
\begin{align}
	X_{(0)}^{\ua} = (0\,,0\,,0\,,0\,,0\,,-\ell)\,,
\end{align}
which corresponds to the bi-twistor 
\begin{align}
	X_{(0)} = X_{(0)}^{\ua}(\S_{\ua}) = \ell C^{-1}\,. 
\end{align}
Then, any bi-twistor corresponding to a point in AdS$_5$ can be written as 
\begin{align} \label{x bt general}
	\hat{X} = \ell g C^{-1} g^{\T}\,, \qquad g \in \sSU(2,2)\,. 
\end{align}
The group element $g$, however, is not uniquely defined. It is defined modulo 
\begin{align}
	g \sim g \l\,, \qquad \l \in \sUSp(2,2)\,. 
\end{align}
This completes the proof.

\section{$\N$-extended AdS superspace in five dimensions} \label{susy ads}

The supergroup $\sSU(2,2|\N)$ is the isometry group of five-dimensional $\N$-extended AdS superspace $\text{AdS}^{5|8\N}$, eq.  \eqref{n extended ads}.
Here we will describe an embedding formalism for $\text{AdS}^{5|8\N}$ constructed in terms of supertwistors of $\sSU(2,2|\N)$.

\subsection{Algebraic background}

The supergroup $\sSU(2,2|\N)$
naturally acts on the space of {\it even} supertwistors and on
the space of {\it odd} supertwistors.

An arbitrary supertwistor $\bm{T}$ is a column vector
\bea
\bm{T} = \left(\bm{T}_{A}\right) = \left(\begin{array}{c}
	\bm{T}_{\ah} \\
	\hline \hline
	\bm{T}_{i}
\end{array}\right)\,, \qquad i = 1\,, \ldots\,, \N\,. 
\eea
In the case of even supertwistors, $ \bm{T}_\hal$ is bosonic
and $\bm{T}_i$ is fermionic.
In the case of odd supertwistors, $\bm{T}_\hal$ is fermionic while $\bm{T}_i$ is bosonic.
The even and odd supertwistors are called pure.
We introduce the parity function $\e ( \bm{T} )$ defined as:
$\e ( \bm{T} ) = 0$ if $ \bm{T}$ is even, and $\e ( \bm{T} ) =1$ if $\bm T $ is odd.
If we define
\begin{align}
	\e_A= \bigg\{\begin{array}{ccc}
		0 ~,& ~ & A = \ah \\
		1 ~,& ~ & A = i
	\end{array}
\end{align}
then the components $\bm{T}_A$ of a pure supertwistor
 have the following  Grassmann parities
\bea
\e ( \bm{T}_A) = \e ( \bm{T} ) + \e_A \quad (\mbox{mod 2})~.
\eea
The space of even supertwistors is naturally identified with
${\mathbb C}^{4|\cN}$,
while the space of odd supertwistors may be identified with
${\mathbb C}^{\cN |4}$.

Supertwistor space is equipped with the inner product 
\begin{align}
	\braket{\bm{T}}{\bm{S}} = \bm{T}^{\dag}\bm{\O S}\,, \qquad \bm{\O} = \left(\begin{array}{ccc}
		0 & \id_{2} & 0 
		\\
		\id_{2} & 0 & 0 
		\\
		0 & 0 & -\id_{\N}
	\end{array}\right) = \left(\begin{array}{cc}
	\O & 0 \\
	0 & -\id_{\N}
\end{array}\right)\,. 
\end{align}
The inner product is invariant under the supergroup $\sSU(2,2|\N)$
spanned by supermatrices of the form 
\begin{align} \label{susy master eq}
	g = (g_{A}{}^{B}) \in \sSL(4|\N)\,, \qquad g^{\dag}\bm{\O}g &= \bm{\O}\,.
\end{align}

Associated with a supertwistor $\bm{T}$ is its dual
\begin{align}
	\bar{\bm{T}} := \bm{T}^{\dag}\bm{\O} = (\bar{\bm{T}}{}^{A}) = (\bar{\bm{T}}{}^{\ah}\,, -\bar{\bm{T}}{}^{i})\,, \qquad \bar{\bm{T}}{}^{i} := (\bm{T}_{i})^{*}\,. 
\end{align}
$\sSU(2,2|\N)$ acts on supertwistors and their duals as 
\begin{align}
	\bm{T}_{A} \rightarrow g_{A}{}^{B}\bm{T}_{B}\,, \qquad \bar{\bm{T}}{}^{A} \rightarrow \bar{\bm{T}}{}^{B}(g^{-1})_{B}{}^{A}\,. 
\end{align}

The superalgebra $\mathfrak{su}(2,2|\N)$ is spanned by elements of the form
\bsubeq \label{superalg reparam}
\begin{align} 
	\mathfrak{g} = \left(\begin{array}{c|c||c}
		\o_{\a}{}^{\b} - (\frac{1}{2}\o^{45} - \frac{\ri\N}{\N-4}\t) \d_{\a}{}^{\b} & - \frac{\ri}{2}(\o^{a4} + \o^{a5})(\s_{a})_{\a\bd} & 2\psi_{\a}{}^{j} \\
		\hline 
		\frac{\ri}{2}(\o^{a4} - \o^{a5})(\tilde{\s}_{a})^{\ad\b} & 
		-\bar{\o}^{\ad}{}_{\bd} + (\frac{1}{2}\o^{45} + \frac{\ri\N}{\N-4}\t)\d^{\ad}{}_{\bd} & 2\bar{\e}^{\ad j}
		\\
		\hline \hline 
		2\e_{i}{}^{\b} & 2\bar{\psi}_{i\bd} & \frac{4\ri}{\N-4}\t\d_{i}{}^{j} + \L_{i}{}^{j}
	\end{array}\right)\,,
\end{align}
with 
\begin{align}
	\o_{\a}{}^{\b} = \frac{1}{2}\o^{ab}(\s_{ab})_{\a}{}^{\b}\,, \qquad  
	\o^{\ua\ub} = - \o^{\ub\ua} \,, \t \in \mathbb{R}\,, \qquad \L = (\L_i{}^j) \in \mathfrak{su}(\N)\,. 
\end{align}
\esubeq
The elements of $\mathfrak{su}(2,2|\N)$ satisfy the conditions
\begin{align}
	\text{str} \, \mathfrak{g} = 0\,, \qquad \mathfrak{g}^{\dag}\bm{\O} + \bm{\O}\mathfrak{g}= 0\,. 
\end{align}
More details on the AdS superalgebra will be elaborated in section \ref{geometry}. 

\subsection{Supertwistor realisation}

In complete analogy with the bosonic construction described in section \ref{boson ads}, consider four {\it even} supertwistors with linearly independent bodies\footnote{The terminology ``body'' and ``soul'' follows \cite{DeWitt}.}
\begin{align}
\bm{T}^{\tmu}	=(\bm{T}_{A}{}^{\tmu})\,, \qquad \e (\bm{T}_{A}{}^{\tmu}) = \e_A\, , \qquad
\tmu = 1\,,2\,,3\,,4\,.
\end{align}
We restrict our attention to those quartets satisfying the constraint 
\begin{align} \label{susy const}
	\bar{\bm{T}}{}^{\tmu A} \bm{T}_{A}{}^{\tnu} = \O^{\tmu\tnu}\,, \qquad 
	\bar{\bm{T}}{}^{\tmu A} :=  (\bm{T}_{B}{}^\tmu)^* \,\bm{\O}^{BA}\,.
\end{align}
The space of such quartets will be denoted
$\mathfrak{F}_{\N}$.
On $\mathfrak{F}_{\N}$, we introduce the equivalence relation
\bsubeq \label{susy equiv}
\begin{align} 
	\bm{T}_{A}{}^{\tmu} &\sim \bm{T}_{A}{}^{\tnu} \l_{\tnu}{}^{\tmu} \,, \qquad \l \in \sUSp(2,2)\,, \label{susy usp}
	\\
	\bm{T}_{A}{}^{\tmu} &\sim \re^{\ri\vf}\bm{T}_{A}{}^{\tmu} \,, \qquad \vf \in \mathbb{R}\,. \label{susy u1}
\end{align}
\esubeq
We emphasise that the condition \eqref{susy const} is invariant under the equivalence transformations \eqref{susy equiv}. 
For $\N > 0$, the constraint \eqref{det const} is substituted with 
\eqref{susy u1}.

The supergroup $\sSU(2,2|\N)$ acts on $\mathfrak{F}_{\N}$ as 
\begin{align}
	\bm{T}_{A}{}^{\tmu} \rightarrow g_{A}{}^{B}\bm{T}_{B}{}^{\tmu}\,, \qquad g \in \sSU(2,2|\N)\,. 
\end{align}
This action is naturally extended to the quotient space $\mathfrak{F}_{\N} / \sim$, which proves to be a homogeneous space of $\sSU(2,2|\N)$. 
To show this, we note that the quartets $\bm{T}_{A}{}^{\tmu}$ can be embedded into a group element 
\bea
g = ( \bm{T}_{A}{}^{\tmu} , \bm{\X}_{A}{}^{\ti})  \in \sSU(2,2|\N)~, \qquad  \ti =1,\dots, \cN\, , 
\eea
and $\sSU(2,2|\N)$ acts transitively on itself from the left. 
It turns out that 
\begin{align} \label{ads susy def}
	\mathfrak{F}_{\N} / \sim~ =~ \text{AdS}^{5|8\N}\,. 
\end{align}

\subsection{Stabiliser}

To prove \eqref{ads susy def}, it suffices to choose a base point and determine its stabiliser. 
As a base point, $\bm{T}^{(0)}$, we choose 
\begin{align} \label{origin}
	\bm{T}^{(0)} = \left(\begin{array}{c}
		\id_{4} 
		\\
		\hline 
		\hline 
		0
	\end{array}\right)\,.
\end{align}
This is the simplest quartet satisfying the constraint \eqref{susy const}.
By definition, the stabiliser $H$ of $\bm{T}^{(0)}$ consists of those elements $h \in \sSU(2,2|\N)$ that satisfy the condition 
\begin{align}
	h \bm{T}^{(0)} = \left(\begin{array}{c}
		\re^{\ri\vf}\l 
		\\
		\hline 
		\hline 
		0
	\end{array}\right)\,, \qquad \vf \in \mathbb{R}\,, \quad \l \in \sUSp(2,2)\,. 
\end{align}
That is, it consists of those group elements that map $\bm{T}^{(0)}$ to an equivalent point with respect to \eqref{susy equiv}. 
These conditions imply that the group elements $h$ take the form
\bsubeq
\begin{align} \label{little group element}
	h &=
	 \left(\begin{array}{c||c}
		~\re^{\ri\vf} \id_{4}~ & 0
		\\
		\hline \hline 
		0 & ~\re^{\frac{4}{\N}\ri\vf}\id_{\N}~
	\end{array}\right) 
\left(
\begin{array}{c||c}
	~\l~ & 0 
	\\
	\hline \hline 
	0 & ~\id_{\N}~
\end{array}
\right)
\left(
\begin{array}{c||c}
	~\id_{4}~ & 0 
	\\
	\hline \hline 
	0 & ~U~
\end{array}
\right)\,,
\end{align}
with 
\begin{align}
	\vf \in \mathbb{R}\,, \qquad \l & \in \sUSp(2,2)\,, \qquad U \in \sSU(\N)\,. 
\end{align}
\esubeq
We therefore have $H$ is isomoprhic to
\begin{align}
	\sUSp(2,2)\times\sU(\N)\,. 
\end{align}

\subsection{Bi-supertwistor realisation}

In this subsection we will detail a bi-supertwistor realisation of the AdS superspace described above. 
Given a point in $\mathfrak{F}_{\N}$, we associate with it the following graded antisymmetric supermatrix 
\begin{align} \label{sbt def}
	\bm{X}_{AB} := \ell \mathfrak{C}_{\tmu \tnu} \bm{T}_{A}{}^{\tmu} \bm{T}_{B}{}^{\tnu} = - (-1)^{\e_A \e_B}\bm{X}_{BA}\,. 
\end{align}
These supermatrices are invariant under arbitrary equivalence transformations of the form \eqref{susy usp}. 
Because of \eqref{susy u1}, they are defined modulo the equivalence relation 
\begin{align} \label{sbt u1}
	\bm{X}_{AB} \sim \re^{\ri\vf}\bm{X}_{AB}\,, \qquad \vf \in \mathbb{R}\,,
\end{align}
and hence can be used to parametrise the space $\mathfrak{F}_{\N} / \sim$. 
Using the dual supertwistors we can introduce the following supermatrix 
\begin{align} \label{dual sbt def}
	\bm{\bar{X}} := \bm{\O}\bm{X}^{\dag}\bm{\O}\,, \qquad \bm{\bar{X}}{}^{AB} = -\ell \mathfrak{C}_{\tmu \tnu}  \bm{\bar{T}}{}^{\tmu A}\bm{\bar{T}}{}^{\tnu B}\,. 
\end{align}
The important properties of the supermatrices $\bm{X}$ and $\bm{\bar{X}}$ are
\bsubeq \label{sbt properties}
\begin{align}
	\bm{X}_{[AB}\bm{X}_{CD}\bm{X}_{E\}F} &= 0 \implies \bm{X}_{[AB}\bm{X}_{CD}\bm{X}_{EF\}} = 0\,, \label{graded antisym sbt}
	\\
	\bm{\bar{X}}{}^{AB}\bm{X}_{BA} &= 4\ell^{2}\,, \label{sbt l squared}
	\\
	(-1)^{\e_{C}}\bm{X}_{AC}\bar{\bm{X}}{}^{CD}\bm{X}_{DB} &= \ell^{2}\bm{X}_{AB}\,, \label{triple product sbt}
\end{align}
\esubeq
where $[\ldots\}$ denotes the graded antisymmetrisation of indices.
Property \eqref{graded antisym sbt} follows from the fact that $\bm{X}_{AB}$ is constructed from four supertwistors. 
An important implication of this property is that the body of the bosonic block $\bm{X}_{ij}$ defined by
\begin{align}
	\bm{X} = \left(\begin{array}{c||c}
		\bm{X}_{\ah\bh} & \bm{X}_{\ah j}
		\\
		\hline \hline
		\bm{X}_{i\bh} & \bm{X}_{ij}
	\end{array}\right)\,,
\end{align}
vanishes. Indeed, let us consider the case 
\begin{align}
	\bm{X}_{(ij}\bm{X}_{kl}\bm{X}_{mn)} = 0\,. 
\end{align}
Since $\bm{X}_{ij}$ is symmetric, the only non-zero solution to the above is a symmetric, bodiless matrix $\bm{X}_{ij}$.

Now, we would like to describe the superspace AdS$^{5|8\N}$ solely in terms of bi-supertwistors, without any reference to the supertwistor realisation above. 
Let us consider the space of graded antisymmetric supermatrices $\bm{X}_{AB}$, 
\begin{subequations}
\begin{align}
	\bm{X}_{AB} &= -(-1)^{\e_A\e_B}\bm{X}_{BA}\, , 
	\\
	\e(\bm{X}_{AB}) &= \e_{A}+ \e_{B}\,. 
\end{align}
\end{subequations}
In this space, we consider the surface $\mathfrak{L}$ consisting of those supermatrices which obey the constraints \eqref{sbt properties}.
We then introduce the quotient space $\mathfrak{L}/\sim$, where
the equivalence relation is given by \eqref{sbt u1}. 
Our goal is to show that $\mathfrak{L}/\sim$ can be identified with AdS$^{5|8\N}$. 

We begin by considering all the elements of $\mathfrak{L}$ with vanishing soul.
These elements take the form 
\begin{align} \label{3.27}
	\bm{X}_{AB}| := X_{AB} = \left(\begin{array}{c||c}
		X_{\ah\bh} & 0
		\\
		\hline\hline
		0 & ~0~
	\end{array}\right)\,, \qquad X_{\ah\bh} = -X_{\bh\ah} \in {\mathbb C}\,, 
\end{align}
where $\bm{X}_{AB}|$ denotes the soulless component of $\bm{X}_{AB}\,.$
The space of such elements will be denoted $\mathfrak{L}|$. 

It turns out that $\mathfrak{L}|/\sim $ is a homogeneous space of the subgroup of $\sSU(2,2|\N)$ spanned by matrices of the form
\begin{align} \label{reduced supergroup element}
	v = \left(\begin{array}{c||c}
	\re^{\ri\vf} \S & ~0~ \\
	\hline \hline 
	0 & \re^{\frac{4\ri\vf}{\N}} U
\end{array}\right)\,, \qquad \vf \in \mathbb{R}\,, \quad \S \in \sSU(2,2)\,, \quad  U \in \sSU(\N)\,. 
\end{align} 
The proof is analogous to that described in subsection \ref{bt realisation} and is described below. 
As a base point we can choose
\begin{align} \label{pref Y}
	X_{(0)} = \left(\begin{array}{c||c}
		\ell C^{-1} & ~0~ \\
		\hline \hline 
		0 & 0
	\end{array}\right)\,. 
\end{align}
Then, any point in $\mathfrak{L}|/\sim $ can be reached by a group transformation of the form 
\begin{align} \label{l y l}
	X = v X_{(0)} v^{\T}\,,
\end{align}
with $v$ given by \eqref{reduced supergroup element}. 
As in the bosonic case, the transformation $v$ is not uniquely defined. 
The stabiliser of the bi-supertwistor $X_{(0)}$ consists precisely of those group elements that take the form \eqref{little group element}. 
We therefore have that $\mathfrak{L}| / \sim$ coincides with the body of the coset superspace $\sSU(2,2|\N) / \left(\sUSp(2,2) \times \sU(\N)\right)\,.$

To prove \eqref{l y l}, we note that  $X_{\ah\bh}$ in \eqref{3.27} can be written as
\begin{align}
	X_{\ah\bh} = X^{\ua}(\S_{\ua})_{\ah\bh}\,. 
\end{align}
The matrix $\hat{X} = (X_{\ah\bh})$ is non-singular. 
Because of property \eqref{triple product sbt}, we have 
\begin{align} \label{ybary}
	\hat{X}^{\dag} \O \hat{X} = \ell^{2}\O \implies \bar{X}^{\ah\gh}X_{\gh\bh} = \ell^{2}\d^{\ah}{}_{\bh}\,. 
\end{align}
This property has two implications: 
\begin{enumerate}
\item
the mutually conjugate six-vectors $\bar{X}^{\ua}$ and $X^{\ua}$ satisfy the following
\bsubeq
\begin{align} 
	\bar{X}^{\ua}X_{\ua} = -\ell^{2}\,,
	\label{i}
\end{align}
\item
$\bar{X}^{\ua}$ and $X^{\ua}$  are linearly dependent, which follows from
\begin{align}
	\bar{X}^{\ua}X^{\ub}(\tilde{\S}_{\ua\ub})^{\ah}{}_{\bh} = 0\,,
	\label{ii}
\end{align}
\esubeq
\end{enumerate}
where we have used \eqref{i}. 
Property \eqref{i} can be seen by taking the trace of \eqref{ybary}.
To prove property \eqref{ii} it is instructive to write out the above expression explicitly,
\begin{align}
	\bar{X}^{\ua}X^{\ub}(\tilde{\S}_{\ua\ub})^{\ah}{}_{\bh} 
	&= 
	-\frac{1}{4}\bar{X}^{\ua}X^{\ub}\left(\tilde{\S}_{\ua}\S_{\ub} - \tilde{\S}_{\ub}\S_{\ua}\right)^{\ah}{}_{\bh}\,,
\end{align}
and then make use of \eqref{s stilde clifford}. 
Then we have 
\begin{align}
	X^{\ua} = \re^{\ri\vf}Y^{\ua}\,, \qquad \vf \in \mathbb{R}\,, \quad Y^{\ua} \in \text{AdS}_{5}\,. 
\end{align}
Now we can repeat the argument from the non-supersymmetric case.
Specifically, we have 
\begin{align}
	X^{\ua} = \re^{\ri\vf}s(Y)^{\ua}{}_{\ub}Y_{(0)}^{\ub}\,, \qquad s(Y) \in \sSO_{0}(4,2)\,,
\end{align}
for a base point $Y_{(0)}^{\ua}$. 
Expression \eqref{l y l} follows. 

We now note that both the full supergroup, $\sSU(2,2|\N)$, and a generic bi-supertwistor $\bm{X}_{AB} \in \mathfrak{L} / \sim$ have $8\N$ real Grassmann odd degrees of freedom. 
We conclude that, by replacing $v$ in \eqref{l y l} with some $g \in \sSU(2,2|\N)$, we can reach an arbitrary $\bm{X} \in \mathfrak{L} / \sim$ by a transformation acting on $X_{(0)}$. 
That is, the space $\mathfrak{L} / \sim$ is itself a homogeneous space of $\sSU(2,2|\N)$, which shows the equivalence of the bi-supertwistor and supertwistor realisations of AdS$^{5|8\N}$. 

\subsection{Projection-like operator}

There is an additional structure we can introduce, making use of the supertwistor and bi-supertwistor realisations of AdS$^{5|8\N}$. 
Let us introduce the following supermatrix
\begin{align}
\bm{Y} = (\bm{Y}_{A}{}^{B})\, , \qquad
	\bm{Y}_{A}{}^{B} := \bm{T}_{A}{}^{\tmu}\O_{\tmu\tnu}\bar{\bm{T}}{}^{\tnu B}~,
\end{align}
with the transformation law under the AdS supergroup 
\begin{align}
	\bm{Y} \rightarrow g \bm{Y} g^{-1}\,, \qquad 
	g \in \sSU(2,2|\N)\,. 
\end{align}
This structure is unique to the supersymmetric case in the sense that for $\N=0$ it coincides with the unit matrix, 
\bea
\cN=0 \,: \qquad \bm{Y} = {\mathbbm 1}_4\, .
\eea
It is invariant under the equivalence relation \eqref{susy equiv}, and satisfies the following properties: 
\bsubeq
\begin{align}
	\bm{Y}_{A}{}^{C}\bm{Y}_{C}{}^{B} &= \bm{Y}_{A}{}^{B}\,, 
	\\
	\bm{Y}_{A}{}^{B}\bm{T}_{B}{}^{\tmu} &= \bm{T}_{A}{}^{\tmu}
	\\
	\bar{\bm{T}}{}^{\tmu B} \bm{Y}_{B}{}^{A}&= \bar{\bm{T}}{}^{\tmu A} \,,
	\\
	\bm{Y}_{A}{}^{C}\bm{X}_{CB} &= \bm{X}_{AB}\,, 
	\\
	\bar{\bm{X}}^{AC}	\bm{Y}_{C}{}^{B} &= \bar{\bm{X}}^{AB}\,, 
	\\
	(-1)^{\e_{A}}\bm{Y}_{A}{}^{A} &= 4\,. 
\end{align}
\esubeq
The operator $\bm{Y}_{A}{}^{B}$ can be related to the bi-supertwistors $\bm{X}_{AB}$ as 
\begin{align}
	\bm{Y}_{A}{}^{B} = \frac{(-1)^{\e_{C}}}{\ell^{2}}\bm{X}_{AC}\bar{\bm{X}}{}^{CB}\,. 
\end{align}

A similar structure is available in the case of $3$D $(p,q)$ AdS superspaces, see \cite{Kuzenko:2023yak}.

\subsection{$\sSU(2,2|\N)$-invariant two-point function}

A powerful application of the formalism described above is the construction of manifestly invariant two-point functions. 
Indeed, let $\bm{T}$ and $\bm{T}'$ be arbitrary points of $\mathfrak{F}_{\N}$. 
The following two-point function 
\begin{align} \label{tp function}
	\o(\bm{T},\bm{T}') := -\frac{1}{4}\ell^{2}\mathfrak{C}_{\tmu\tnu}\mathfrak{C}_{\tro\tsi}\braket{\bm{T}^{\tmu}}{\bm{T}'^{\tro}}\braket{\bm{T}^{\tnu}}{\bm{T}'^{\tsi}}
\end{align}
is $\sSU(2,2|\N)$-invariant. It is also invariant under the equivalence transformations \eqref{susy equiv}, and is therefore defined on the quotient space \eqref{ads susy def}.

The two-point function \eqref{tp function} can be expressed in terms of the bi-supertwistors $\bm{X}$ and $\bar{\bm{X}}$. 
Let us denote the bi-supertwistors corresponding to $\bm{T}$ and $\bm{T}'$ as $\bm{X}$ and $\bm{X}'$. Then the two-point function takes the form 
\begin{align} \label{sbt tp function}
	\o(\bm{X},\bm{X}') = - \frac{1}{4}\bar{\bm{X}}^{AB}\bm{X}'_{BA}\,. 
\end{align}
Making use of the operator $\bm{Y}$ introduced above, we can also consider the following $n$-point function 
\begin{align} \label{y n point}
	\x(\bm{Y}_{1}\,, \ldots\,, \bm{Y}_{n}) := \frac{1}{4}\text{Str}\left(\bm{Y}_{1}\bm{Y}_{2}\ldots\bm{Y}_{n}\right)\,. 
\end{align}
In the non-supersymmetric case, $\N = 0$, \eqref{tp function} and \eqref{sbt tp function} coincide with the AdS$_5$ two-point function $X^{\ua}X'_{\ua}$, and \eqref{y n point} is constant, $\x(Y_{1}\,, \ldots Y_{n}) = 1\,.$ 

Recently, a similar construction to the above was used to describe a worldline superparticle model in AdS$^{4|4\N}$, see \cite{Koning:2023ruq, KKR2}.
In the present context, such a model is described by the following action 
\begin{align} \label{superparticle model}
	S = \frac{1}{2}\int \text{d}\t \mathfrak{e}^{-1}\left\{\o(\dot{\bm{X}},\dot{\bm{X}}) + \a \,\x(\dot{\bm{Y}},\dot{\bm{Y}}) - 
	(\mathfrak{e}m)^{2} \right\}\,, 
\end{align}
for a real parameter $\a$. 
It can be shown that the $\a$-term vanishes when the Grassmann variables are switched off, and the resulting model coincides with the bosonic one. 

\section{Coset representative} \label{coset rep section}

A key role in the study of the geometry of a coset space is played by the coset representative. 
A global coset representative for a homogeneous space $\cM = G / H_{x_{0}}$, where $G$ is a Lie group and $H_{x_{0}}$ is the stabiliser (also known as the isotropy group or little group) of some point $x_{0} \in \cM$, is an injective map $\cS: \cM \rightarrow G$ such that $\p \circ \cS = \text{id}_{\cM}$, where $\p$ denotes the canonical projection $\p: G \rightarrow G/H_{x_{0}}$.  
For many homogeneous spaces, a global coset representative does not exist. 
In such cases, local coset representatives $\cS_{A}: U_{A} \rightarrow G$ with the property $\p \circ \cS_{A} = \text{id}_{U_{A}}$ can be defined on open charts $\{U_{A}\}$ that provide an atlas for $\cM$. 
In the intersection of two charts, $U_{A}$ and $U_{B}$, $U_{A} \cap U_{B} \neq \emptyset$, the corresponding coset representatives $\cS_{A}$ and $\cS_{B}$ are related by a little group transformation, $\cS_{B}(x) = \cS_{A}(x)h_{AB}(x)\,,$ with $h_{AB}(x) \in H_{x_{0}}$.

A global coset representative exists for $d$-dimensional AdS in the context of the standard embedding formalism, see \cite{Kuzenko:1995aq}. 
For the embedding formalism described in this work, suitable for AdS superspace, an exponential coset parametrisation is used. 
In this case we find local coset representatives and their associated local coordinate systems. 
These are detailed below. 

\subsection{AdS space ($\N = 0$)}

The Lie algebras $\mathfrak{su}(2,2)$ and $\mathfrak{usp}(2,2)$ are given by expressions \eqref{su(2,2) generic algebra element} and  \eqref{so(4,1) generic algebra element}, respectively.
Let $\mathfrak{K}$ be the complement of $\mathfrak{usp}(2,2)$ in $\mathfrak{su}(2,2)$.
We then have
\begin{align} \label{su22 decomp}
	\mathfrak{su}(2,2) = \mathfrak{usp}(2,2) \oplus \mathfrak{K}\,.
\end{align}
The elements of $\mathfrak{K}$ take the form 
\begin{align} \label{compel}
	\mathfrak{t} = \left(\begin{array}{c|c}
		- \frac{1}{2}\o^{45}\d_{\a}{}^{\b} & - \frac{\ri}{2}(\o^{a 5})(\s_{a})_{\a\bd}
		\\
		\hline 
		- \frac{\ri}{2}(\o^{a5})(\tilde{\s}^{a})^{\ad\b} & \frac{1}{2}\o^{45}\d^{\ad}{}_{\bd}
	\end{array}\right) \in \mathfrak{K}\,, \qquad \o^{\ha 5} \in \mathbb{R}\,.
\end{align}
Then, for a group element $g \in \sSU(2,2)$ in a neighbourhood of $\id_{4}$, it holds that 
\begin{align}
	g = \re^{\mathfrak{t} + \mathfrak{h}}\,, \qquad \mathfrak{h} \in \mathfrak{usp}(2,2)\,. 
\end{align}
We seek to (locally) factorise $g$ in the following way 
\begin{align}
	g = \cS(y) h\,, \qquad \cS(y) \in \sSU(2,2)\, ,
	\qquad h \in \sUSp(2,2)\,,
\end{align}
where $\cS(y) $ is the coset representative, and $y$ are local coordinates on AdS$_5$. 
We can make the following ansatz for $\cS(y)$:
\begin{align}\label{bosonic coset rep}
	\cS(y) = \left(
	\begin{array}{c|c}
		~\x\r \id_{2}~ & -\ri x \\
		\hline
		-\ri \tilde{x} & \x\r^{-1} \id_{2}
	\end{array}
	\right) = \re^{\mathfrak t}
	\,, 
\end{align}
with 
\bsubeq
\begin{align}
	x &:= x^{a}(\s_{a}) = x^{\dag}\,, \qquad \tilde{x}:= x^{a}(\tilde{\s}_{a}) = \tilde{x}^{\dag}\,,
	\\
	\x &:= (1+x^{2})^{\frac{1}{2}}\,, \qquad \,~~x^{2} = x^{a}x^{b}\eta_{ab}\,,
	\\
	\quad \qquad \r &= \bar{\r}\,.  
\end{align}
\esubeq
It follows that the local coordinates $x^{a}$ and $\r$ are real, and the coordinate chart is specified by 
\begin{align}
	x^{2} > -1\,, \qquad \r \neq 0\,. 
\end{align}
The bi-twistor corresponding to \eqref{bosonic coset rep} is given by 
\bsubeq
\begin{align} \label{cr bt}
	X_{\ah\bh} &= \ell \mathfrak{C}_{\tmu\tnu}\cS(y)_{\ah}{}^{\tmu}\cS(y)_{\bh}{}^{\tnu} = 
	\ell \left(
	\begin{array}{c|c}
		\big(\x^{2}\r^{2} + x^{2}\big)\ve_{\a\b} & - \ri\x(\r + \r^{-1})x_{\a}{}^{\bd}
		\\
		\hline 
		\ri\x(\r+\r^{-1})x^{\ad}{}_{\b}	& - \big(\x^{2}\r^{-2} + x^{2}\big)\ve^{\ad\bd} 
	\end{array}
	\right)\,,
\end{align}
where
\begin{align}
	\cS(y)_{\ah}{}^{\tmu} &:= \cS(y)_{\ah}{}^{\bh}T^{(0)}_{\bh}{}^{\tmu} \,, \qquad 
	T^{(0)} = \id_{4}\,. 
\end{align}
\esubeq
Associated with $X_{\ah\bh}$ is the following point in AdS$_5$
\begin{align}
	X^{\ua} = \left(
	\ell \x (\r + \r^{-1}) x^{a} \,, - \frac{1}{2}\ell \x^{2} (\r^{2} - \r^{-2})\,,
	-\ell\big(x^{2} + \frac{1}{2}\x^{2}(\r^{2} + \r^{-2})\big)
	\right)\,. 
\end{align}
Instead of the coset parametrisation used above, we could choose a coset representative corresponding to Poincar\'e coordinates in AdS$_5$. 
This coset representative, used in \cite{Claus:1999zh, Kuzenko:2007aj}, is given by
\begin{align}
	\cS(y) = \left(
	\begin{array}{c|c}
		\r^{\frac{1}{2}}\id_{2} & ~0~ 
		\\
		\hline 
		-\ri\r^{\frac{1}{2}}\tilde{x} & \r^{-\frac{1}{2}}\id_{2}
	\end{array}
	\right)\,.
\end{align}
Its corresponding bi-twistor is 
\begin{align} \label{pp bt}
	X_{\ah\bh} = \frac{1}{z}\left(
	\begin{array}{c|c}
		~\ve_{\a\b}~ & -\ri x_{\a}{}^{\bd}
		\\
		\hline
		\ri x^{\ad}{}_{\b} & -\big((\ell z)^{2} + x^{2}\big)\ve^{\ad\bd}
	\end{array}
	\right)\,, \qquad z:= (\ell \r)^{-1}\,. 
\end{align}
The real coordinates $z>0$ and $x^{a}$ parametrise AdS$_5$ in the Poincar\'e patch.
They are related to the embedding coordinates $X^{\ua}$ as follows
\begin{align}
	X^{\ua} = \frac{1}{z}\left(
	x^{a} \,, -\frac{1}{2}(1 - x^{2} - (\ell z)^{2})\,,
	-\frac{1}{2}(1 + x^{2} + (\ell z)^{2} )
	\right)\,. 
\end{align}

\subsection{$\N \neq 0$} \label{section 4.2}

The analysis of the previous section can be extended to the supersymmetric case in similar fashion. 
Coset representatives for the $\N = 1$ superspaces AdS$^{5|8} \times S^{1}$ and AdS$^{5|8}$ are given in \cite{Kuzenko:2001ag,Kuzenko:2007aj}. 
Building on the approach of these references, again denoting the local coordinates by $y$, we choose the coset representative in a supersymmetric generalisation of Poincar\'e coordinates as
\begin{align} \label{pp coset rep}
	\cS(y) &= g(\bm{z})\cdot g_{S} \cdot g_{D}
	\\ 
	\notag
	&= 
	\left(
	\begin{array}{c|c||c}
		\id_{2} & ~0~ & ~0~ \\
		\hline
		-\ri \tilde{x}_{+} & \id_{2} & 2\bar{\q} \\
		\hline \hline
		2 \q & 0 & \id_{\N}
	\end{array}
	\right)
	\left(
	\begin{array}{c|c||c}
		\id_{2} & 2\eta \bar{\eta} & 2 \eta
		\\ \hline 
		0 & \id_{2} & 0 
		\\
		\hline \hline
		~0~ & 2\bar{\eta} & \id_{\N}
	\end{array}
	\right)
	\left(
	\begin{array}{c|c||c}
		\r^{\frac{1}{2}}\id_{2}& 0 & 0 \\
		\hline 
		0 & \r^{-\frac{1}{2}}\id_{2} & 0 \\
		\hline \hline 
		0 & 0 & \id_{\N}
	\end{array}
	\right)
	\\
	\notag 
	&= 
	\left(
	\begin{array}{c|c||c}
		\r^{\frac{1}{2}}\d_{\a}{}^{\b} 
		& 
		2 \r^{-\frac{1}{2}}\eta_{\a}{}^{k}\bar{\eta}_{k \bd} 
		& 
		2 \eta_{\a}{}^{j}
		\\
		\hline 
		~-\ri \r^{\frac{1}{2}}x_{+}^{\ad\b} ~
		&
		\r^{-\frac{1}{2}}\big(\d^{\ad}{}_{\bd} - 2\ri x_{+}^{\ad\g}\eta_{\g}{}^{k}\bar{\eta}_{k \bd} + 4\bar{\q}^{\ad k}\bar{\eta}_{k \bd}\big)
		&
		2 \big( \bar{\q}^{\ad j} - \ri x_{+}^{\ad\g}\eta_{\g}{}^{j} \big)
		\\
		\hline \hline
		2\r^{\frac{1}{2}}\q_{i}{}^{\b} 
		& 
		2\r^{-\frac{1}{2}}\big(\bar{\eta}_{i\bd} + 2 \q_{i}{}^{\g}\eta_{\g}{}^{k}\bar{\eta}_{k\bd} \big) 
		&
		\d_{i}{}^{j} + 4\q_{i}{}^{\g}\eta_{\g}{}^{j} 
	\end{array}
	\right)\,,
\end{align}
where $x_{\pm}^{a} = x^{a} \pm \ri\q_{i}\s^{a}\bar{\q}^{i}$.
The coset representative $g(\bm{z})$ corresponds to that of $4$D $\N$-extended Minkowski superspace. 
This coordinate system will be referred to as Poincar\'e-like. 

The action of $\cS(y)$ on the base point $\bm{T}^{(0)}$, see \eqref{origin}, yields the general form of the quartet $\bm{T}_{A}{}^{\tmu}$ in Poincar\'e-like coordinates 
\begin{align}
	\bm{T}_{A}{}^{\tmu} = 
	\left(
	\begin{array}{c|c}
		~\r^{\frac{1}{2}}\d_{\a}{}^{\b} ~
		& 
		2 \r^{-\frac{1}{2}}\eta_{\a}{}^{k}\bar{\eta}_{k \bd} 
		\\
		\hline 
		~-\ri \r^{\frac{1}{2}}x_{+}^{\ad\b} ~
		&
		\r^{-\frac{1}{2}}\big(\d^{\ad}{}_{\bd} - 2\ri x_{+}^{\ad\g}\eta_{\g}{}^{k}\bar{\eta}_{k \bd} + 4\bar{\q}^{\ad k}\bar{\eta}_{k \bd}\big)
		\\
		\hline \hline
		2\r^{\frac{1}{2}}\q_{i}{}^{\b} 
		& 
		2\r^{-\frac{1}{2}}\big(\bar{\eta}_{i\bd} + 2 \q_{i}{}^{\g}\eta_{\g}{}^{k}\bar{\eta}_{k\bd} \big) 
	\end{array}
	\right)\,.
\end{align}
Furthermore, the bi-supertwistor $\bm{X}_{AB}$ is 
\begin{align} \label{pp bst}
	\bm{X}_{AB} = \left(\begin{array}{c|c||c}
		\bm{X}_{\ab} & \bm{X}_{\a}{}^{\bd} & \bm{X}_{\a j} \\
		\hline
		\bm{X}^{\ad}{}_{\b} & \bm{X}^{\ad\bd} & \bm{X}^{\ad}{}_{j} \\
		\hline \hline
		\bm{X}_{i\b} & \bm{X}_{i}{}^{\bd} & \bm{X}_{ij}
	\end{array}\right)\,,
\end{align}
where the components are given by 
\bsubeq
\begin{align}
	\bm{X}_{\ab} &= \frac{1}{z} \big(\e_{\a\b} + 4(\ell z)^{2}\eta_{\a}{}^{k}\eta_{\b}{}^{l}\bar{\eta}_{k \gd}\bar{\eta}_{l}{}^{\gd} \big)\,, 
	\\
	\bm{X}_{\a}{}^{\bd} &= \frac{1}{z} \big( 
	-\ri  x_{+}{}_{\a}{}^{\bd}
	+2(\ell z)^{2}\eta_{\a}{}^{k}\bar{\eta}_{k}{}^{\gd} 
	\big( \d^{\bd}{}_{\gd} - 2\ri x_{+}^{\bd \g}\eta_{\g}{}^{l}\bar{\eta}_{l \gd} + 4\bar{\q}^{\bd l}\bar{\eta}_{l\gd}
	\big)
	\big)
	\,,
	\\
	\bm{X}_{\a j} &= 
	\frac{1}{z} \big(2 \q_{\a j} - 4(\ell z)^{2}\eta_{\a}{}^{k}\bar{\eta}_{k\gd}\big(\bar{\eta}_{j}{}^{\gd} + 2\q_{j}{}^{\g}\eta_{\g}{}^{l}\bar{\eta}_{l}{}^{\gd}
	\big) \big) \,,
	\\
	\bm{X}^{\ad\bd} &= \frac{1}{z}\big(
	- (\ell z)^{2}\big(\d^{\ad}{}_{\gd} - 2\ri x_{+}^{\ad\g}\eta_{\g}{}^{k}\bar{\eta}_{\k \gd} + 4\bar{\q}^{\ad k}\bar{\eta}_{k \gd} \big)
	\big(\ve^{\gd\bd} - 2\ri x_{+}^{\bd \s} \eta_{\s}{}^{l}\bar{\eta}_{l}{}^{\gd} + 4\bar{\q}^{\bd l}\bar{\eta}_{l}{}^{\gd}
	\big)
	\\
	\notag 
	& \quad 
	- x_{+}^{2}\ve^{\ad\bd} \big)\,,
	\\
	\bm{X}^{\ad}{}_{j} &= \frac{1}{z} \big(
	-2(\ell z)^{2}
	\big(
	\d^{\ad}{}_{\gd} - 2\ri x_{+}^{\ad\g}\eta_{\g}{}^{k}\bar{\eta}_{k\gd} + 4\bar{\q}^{\ad k}\bar{\eta}_{k \gd}
	\big)
	\big(
	\bar{\eta}_{j}{}^{\gd} + 2\q_{j}{}^{\g}\eta_{\g}{}^{l}\bar{\eta}_{l}{}^{\gd}
	\big)
	\\
	\notag 
	& \quad - 2\ri x_{+}^{\ad\g}\q_{\g j} \big)\,,
	\\
	\bm{X}_{ij} &= \frac{1}{z} \big( 
	4 \q_{i}{}^{\a}\q_{\a j} 
	- 4(\ell z)^{2}\big(
	\bar{\eta}_{\i \ad} + 2\q_{i}{}^{\g}\eta_{\g}{}^{k}\bar{\eta}_{k \ad}
	\big)
	\big(
	\bar{\eta}_{j}{}^{\ad} + 2\q_{j}{}^{\d}\eta_{\d}{}^{l}\bar{\eta}_{l}{}^{\ad}
	\big)
	\big)\,.
\end{align}
\esubeq
The coordinates $z$ and $\r$ are related as in \eqref{pp bt}. 

Using the above formulation we can readily describe the conformal boundary of AdS superspace.
To do so, we should switch from $\bm{X}_{AB} $ to $ z\bm{X}_{AB}$ and take the limit $z \rightarrow 0$.  
In this limit, the Grassmann variables $\eta$ disappear and the resulting bi-supertwistor is given by 
\begin{align}
	\bm{X}^{\text{\tiny{boundary}}}_{AB} = 
	\left(
	\begin{array}{c|c||c}
		\ve_{\ab} 
		& 
		-\ri  x_{+}{}_{\a}{}^{\bd} 
		& 
		2 \q_{\a j}
		\\
		\hline 
		\ri  x_{+}{}^{\ad}{}_{\b}
		&
		- x_{+}^{2} \ve^{\ad\bd} 
		&
		-2 \ri x_{+}{}^{\ad \g}\q_{\g j} 
		\\
		\hline \hline 
		-2 \q_{i \b}
		&
		2 \ri \q_{i \g} x_{+}{}^{\g \bd} 
		&
		4  \q_{i}{}^{\a}\q_{\a j}
	\end{array}
	\right)\,. 
\end{align}
This expression coincides with equation $(3.17)$ in \cite{K-compactified12}.
Making use of the results of \cite{K-compactified12}, $\bm{X}^{\text{\tiny{boundary}}}_{AB}$ can be decomposed into a pair of supertwistors 
\begin{align}
	\bm{T}_{A}{}^{\b} &= \left(\begin{array}{c}
		\d_{\a}{}^{\b}
		\\
		-\ri x_{+}^{\ad\b} 
		\\
		\hline\hline
		2\q_{i}{}^{\b}
	\end{array}\right)\,, \qquad \bm{T}_{A}{}^{\b} \sim \bm{T}_{A}{}^{\g} M_{\g}{}^{\b}\,, \qquad M \in \sGL(2\,,\mathbb{C})\,. 
\end{align}
The local bosonic $(x_{+}^{a})$ and fermionic $(\q_{i}{}^{\a})$ coordinates can be interpreted as coordinates in the chiral subspace of a four-dimensional $\N$-extended Minkowski superspace, $\mathbb{M}^{4|4\N}$.
This is supported by the action of the AdS supergroup, $\sSU(2,2|\N)$, as superconformal transformations on $\mathbb{M}^{4|4\N}$.

An infinitesimal $\sSU(2,2|\N)$ transformation, $g = \id_{4+\N} + \mathfrak{g}$, acts on $\bm{T}_{A}{}^{\b}$ as 
\begin{align}
	\bm{T}_{A}{}^{\b} \rightarrow \bm{T}'_{A}{}^{\b} = \big(\d_{A}{}^{B} + \mathfrak{g}_{A}{}^{B}\big)\bm{T}_{B}{}^{\g} N_{\g}{}^{\b}\,, \qquad N \in \sSL(2,\mathbb{C})\,, 
\end{align}
where $\mathfrak{g}$ is defined in \eqref{superalg reparam}.
One can show that the coordinates $x_{+}^{a}$ and $\q_{i}^{\a}$ transform as 
\bsubeq \label{scf trf}
\begin{align}
	\d \tilde{x}_{+} &= \tilde{a} + (\s + \bar{\s})\tilde{x}_{+} - \bar{\o}\tilde{x}_{+} - \tilde{x}_{+}\o + \tilde{x}_{+}b\tilde{x}_{+} + 4\ri\bar{\e}\q - 4\tilde{x}_{+}\psi\q\,,
	\\
	\d \q &= \e + \frac{1}{\N}\big((\N-2)\s + 2\bar{\s}\big)\q - \q\o + \L\q + \q b\tilde{x}_{+} - \ri\bar{\psi}\tilde{x}_{+} - 4\q\psi\q\,,
\end{align}
\esubeq
where the parameters $\s$, $a^{a}$, and $b^{a}$ are related to those in \eqref{superalg reparam} by the rule
\bsubeq
\begin{align}
	\s &= \frac{1}{2}(\o^{45} - \frac{2\ri\N}{\N-4}\t)\,,
	\\
	a^{a} &= - \frac{1}{2}(\o^{a4} - \o^{a5})\,,
	\\
	b^{a} &= \frac{1}{2}(\o^{a4} + \o^{a5})\,. 
\end{align}
\esubeq
They correspond to 4D Lorentz transformations $(\o\,, \bar{\o})$, translations $(a)$, special conformal transformations $(b)$, Q-supersymmetry $(\e\,, \bar{\e})$ and S-supersymmetry $(\psi\,, \bar{\psi})$ transformations, combined scale and chiral  transformations $(\s)$, and $\sSU(\N)$ transformations $(\L)$. 
The expressions \eqref{scf trf} coincide with the standard superconformal transformations on Minkowski superspace, see, e.g., \cite{Kuzenko:2006mv}.

\section{Superspace geometry} \label{geometry}

In this section we will elaborate on details of the supergeometry of the $\N$-extended AdS superspace introduced above. 
We introduce local bosonic $(x)$ and fermionic $(\q)$ coordinates on AdS superspace as 
\begin{align}
	y^{\hat{M}} = (x^{\hm}\,, \q_{i}{}^{\mh}\,, \bar{\q}^{i}{}_{\mh})\,, \qquad i = 1\,, \ldots \N\,.  
\end{align}

\subsection{Geometric structures in AdS$^{5|8\N}$}

We will begin by formalising some of the earlier discussion about the AdS superalgebra. 
Let us denote the following 
\begin{align}
	\cH := \mathfrak{usp}(2,2) \oplus \mathfrak{u}(1) \oplus \mathfrak{su}(\N)\,. 
\end{align}
Denote the complement of $\cH$ in $\mathfrak{su}(2,2|\N)$ by $\cK$,
\begin{align}
	\mathfrak{su}(2,2|\N) = \cK \oplus \cH\,.
\end{align}
The elements $k \in \cK$ are given by
\begin{align} 
	k = 
	\left(\begin{array}{c|c||c}
		- \frac{1}{2}\o^{45} \d_{\a}{}^{\b} 
		& - \frac{\ri}{2}(\o^{a5})(\s_{a})_{\a\bd}
		& ~2\eta_{\a}{}^{j}~ \\
		\hline 
		-\frac{\ri}{2}(\o^{a5})(\tilde{\s}_{a})^{\ad\b} 
		&  \frac{1}{2}\o^{45}\d^{\ad}{}_{\bd} 
		& 2\bar{\e}^{\ad j}
		\\
		\hline \hline 
		2\e_{i}{}^{\b} 
		& 2\bar{\eta}_{i\bd} 
		& ~0~
	\end{array}\right)\,,
	\qquad \o^{\ha 5} \in \mathbb{R}
	\,. 
\end{align}
The elements $m \in \cH$ are given by  
\bsubeq
\begin{align} \label{stabel}
	m = \left(\begin{array}{c|c||c}
		\frac{1}{2}\o^{ab}(\s_{ab})_{\a}{}^{\b} + \frac{\ri \N}{\N-4}\t \d_{\a}{}^{\b} & - \frac{\ri}{2}(\o^{a4})(\s_{a})_{\a\bd} & 0 \\
		\hline 
		\frac{\ri}{2}(\o^{a4})(\tilde{\s}_{a})^{\ad\b} & \frac{1}{2}\o^{ab}(\tilde{\s}_{ab})^{\ad}{}_{\bd} + \frac{\ri\N}{\N-4}\t\d^{\ad}{}_{\bd} & 0
		\\
		\hline \hline 
		0 & 0 & \frac{4\ri}{\N-4}\t\d_{i}{}^{j} + \L_{i}{}^{j}
	\end{array}\right)\,, 
\end{align}
with 
\begin{align}
	\o^{\ha\hb}\,,\t \in \mathbb{R}\,, \qquad \L \in \mathfrak{su}(\N)\,.
\end{align}
\esubeq

Using the coset representative, one can introduce the Maurer-Cartan form, $\cS^{-1}\text{d}\cS$, which proves to encode all the information about the geometry of our superspace. 
The Maurer-Cartan form can be decomposed into the vielbein and the connection as 
\bsubeq
\begin{align}
	\cS^{-1}\text{d}\cS &= \tb{E} + \bm{\F}\,, 
	\\
	\tb{E} &= (\cS^{-1}\text{d}\cS) |_{\cK}\,,
	\\
	\bm{\F} &= (\cS^{-1}\text{d}\cS)|_{\cH}\,. 
\end{align}
\esubeq
Under a group transformation $g \in \sSU(2,2|\N)$, the coset representative transforms as
\begin{align} \label{group trf on coset rep}
	\cS(y) \rightarrow \cS(y') = g\,\cS(y)\,h^{-1}(g\,,y)\,, \qquad h \in H\,,
\end{align}
and the vielbein and connection transform as 
\bsubeq
\begin{align}
	\tb{E} &\rightarrow h \tb{E} h^{-1}\,,
	\\
	\bm{\F} & \rightarrow h\bm{\F}h^{-1} - \text{d}h\, h^{-1}\,. 
\end{align}
\esubeq

Since the vielbein and the connection are elements of the AdS superalgebra, it is useful to introduce a basis for the superalgebra as
\bsubeq
\begin{align}
	K_{\hA} &:= (P_{\ha}\,, q_{\ah}{}^{i}\,, \bar{q}^{\ah}{}_{i})\,, 
	\\
	H_{\hat{I}} &:= (M_{\ha\hb}\,, \mathbb{J}^{i}{}_{j}\,, \mathbb{Y})\,. 
\end{align}
\esubeq
The basis elements $K_{\hA}$ correspond to supertranslations, while $H_{\hat{I}}$ correspond to Lorentz, $\sSU(\N)$, and $\sU(1)$ transformations.  
They obey the following graded commutation relations 
\bsubeq \label{superalgebra}
\begin{align}
	[M_{\ha\hb}\,, M_{\hc\hd}] &= 
	2\eta_{\hd[\ha} M_{\hb] \hc}
	- 2\eta_{\hc[\ha} M_{\hb] \hd}
	\,,
	\\
	[M_{\ha\hb}\,, P_{\hc}] &= 2\eta_{\hc [\hb}P_{\ha]}
	\,,
	\\
	[P_{\ha}\,, P_{\hb}] &= M_{\ha\hb}
	\,,
	\\
	[\mathbb{J}^{i}{}_{j}\,, \mathbb{J}^{k}{}_{l}] &= 
	\d^{k}{}_{j}\mathbb{J}^{i}{}_{l} - \d^{i}{}_{l}\mathbb{J}^{k}{}_{j}
	\,,
	\\
	[M_{\ha\hb}\,, q_{\ah}{}^{i}] &=
	-(\S_{\ha\hb})_{\ah}{}^{\bh}q_{\bh}{}^{i} 
	\,,
	\\
	[P_{\ha}\,, q_{\ah}{}^{i}] &= 
	\frac{\ri}{2}(\g_{\ha})_{\ah}{}^{\bh}q_{\bh}{}^{i}
	\,,
	\\
	[P_{\ha}\,, \bar{q}^{\ah}{}_{i}] &= \frac{\ri}{2}(\g_{\ha})^{\ah}{}_{\bh}\bar{q}^{\bh}{}_{i}
	\,,
	\\
	[\mathbb{Y}\,,q_{\ah}{}^{i}] &= 
	- q_{\ah}{}^{i}
	\,,
	\\
	\{q_{\ah}{}^{i}\,, \bar{q}^{\bh}{}_{j}\} &= 
	2 \d^{i}{}_{j}M^{\ha\hb}(\S_{\ha\hb})_{\ah}{}^{\bh} 
	+2\ri \d^{i}{}_{j}P^{\ha}(\g_{\ha})_{\ah}{}^{\bh} 
	\\
	\notag
	& \quad 
	- 4\d_{\ah}{}^{\bh}\mathbb{J}^{i}{}_{j} 
	- \frac{\N-4}{\N}\,\d^{i}{}_{j}\d_{\ah}{}^{\bh} \mathbb{Y} 
	\,,
	\\
	[\mathbb{J}^{i}{}_{j}\,, q_{\ah}{}^{k}] &= 
	\d^{k}{}_{j}q_{\ah}{}^{i} - \frac{1}{\N}\,\d^{i}{}_{j}q_{\ah}{}^{k}
	\,,
\end{align} 
\esubeq
with $(\g_{\ha})_{\ah}{}^{\bh}$ defined as \eqref{5d gamma}. 

Any element $n$ of the AdS superalgebra, $n \in \mathfrak{su}(2,2|\N) \,,$ can be written as a linear combination of generators 
\begin{align}
	n &= \frac{1}{2}n^{\ha\hb}M_{\ha\hb} + n_{i}{}^{j}\mathbb{J}^{i}{}_{j} + \ri (n_{\sU(1)}) \mathbb{Y} 
	\\
	\notag 
	& \quad + n^{\ha}P_{\ha} + \ri\big(n_{i}{}^{\ah}q_{\ah}{}^{i} + \bar{n}^{i}{}_{\ah}\bar{q}^{\ah}{}_{i} \big)\,. 
\end{align}
This means that the vielbein can be decomposed as follows 
\begin{align}
	\textbf{E} &= \textbf{E}^{\hA}K_{\hA} = \textbf{E}^{\ha}P_{\ha} + \ri \big(\textbf{E}_{i}{}^{\ah} q_{\ah}{}^{i} + \bar{\textbf{E}}^{i}{}_{\ah}\bar{q}^{\ah}{}_{i}\big)\,,
\end{align}
and the connection can be decomposed as 
\begin{align}
	\bm{\F} &= \bm{\F}^{\hat{I}}H_{\hat{I}} = \frac{1}{2}\bm{\F}^{\ha\hb}M_{\ha\hb} + \bm{\F}_{i}{}^{j}\mathbb{J}^{i}{}_{j} + \ri (\bm{\F}_{\sU(1)})\mathbb{Y}\,. 
\end{align}
In the above expressions we have used the definition
\begin{align}
	\tb{E}^{\hA} = (\tb{E}^{\ha}\,, \tb{E}_{i}{}^{\ah}\,, \bar{\tb{E}}^{i}{}_{\ah})\,. 
\end{align}
The components of the connection, $\bm{\F}^{\hat{I}}\,,$ can be decomposed with respect to the basis $\tb{E}^{\hA}$ to yield the superfields $\F_{\hA}{}^{\hat{I}}$:
\begin{align}
	\bm{\F}^{\hat{I}} = \tb{E}^{\hA}\F_{\hA}{}^{\hat{I}}\,. 
\end{align}

In accordance with the coset construction, we can introduce the curvature and torsion two forms as
\bsubeq \label{torsion and curvature}
\begin{align}
	\tb{R} := \text{d}\bm{\F} - \bm{\F} \wedge \bm{\F}\,, 
	\qquad
	\tb{T} := \text{d}\tb{E} - \tb{E} \wedge \bm{\F} - \bm{\F} \wedge \tb{E}\,. 
\end{align}
The curvature and torsion can alternatively be expressed as
\begin{align}
	\tb{R} = (\tb{E} \wedge \tb{E})|_{\cH}\,, 
	\qquad 
	\tb{T} = (\tb{E} \wedge \tb{E})|_{\cK}\,. 
\end{align}
\esubeq
Under group transformations, \eqref{group trf on coset rep}, they transform covariantly
\begin{align}
	\tb{R}' = h \tb{R} h^{-1}\,, \qquad \tb{T}' = h \tb{T} h^{-1}\,. 
\end{align}
The components of the torsion and curvature two forms are defined via the relations
\bsubeq \label{components def}
\begin{align}
	\tb{T}&= \frac{1}{2}\tb{E}^{\hB}\wedge\tb{E}^{\hA}\big(
	\cT_{\hA\hB}{}^{\hc}P_{\hc} + \ri(\cT_{\hA\hB k}{}^{\gh}q_{\gh}{}^{k} + \cT_{\hA\hB}{}^{k}{}_{\gh}\bar{q}^{\gh}{}_{k})
	\big)
	\,,
	\\	
	\tb{R} &= \frac{1}{2}\tb{E}^{\hB}\wedge\tb{E}^{\hA}\big(
	\frac{1}{2}\cR_{\hA\hB}{}^{\hc\hd}M_{\hc\hd}
	+\cR_{\hA\hB k}{}^{l}\mathbb{J}^{k}{}_{l}
	+\ri(\cR_{\sU(1)})_{\hA\hB}\mathbb{Y}
	\big)\,. 
\end{align}
\esubeq
Building on the approach utilised in \cite{Kuzenko:2007aj, Kuzenko:2023yak, Koning:2023ruq}, using the definitions \eqref{components def} and the graded commutation relations of the generators, the non-vanishing components of the torsion and curvature can be determined. 
They are given by 
\bsubeq \label{ads torsions}
\begin{align}
	\cT^{i}{}_{\ah j}{}^{\bh \ha} &= 2\ri\d^{i}{}_{j}(\g^{\ha})_{\ah}{}^{\bh} 
	\,, \label{global tors}
	\\
	\cT_{\ha}{}^{i}{}_{\ah j}{}^{\bh} &= -\frac{\ri}{2}\d^{i}{}_{j}(\g_{\ha})_{\ah}{}^{\bh}
	\,,
	\\
	\cT_{\ha i}{}^{\ah j}{}_{\bh} &= - \frac{\ri}{2}\d^{j}{}_{i}(\g_{\ha})^{\ah}{}_{\bh}
	\,,
\end{align}
\esubeq
and
\bsubeq \label{ads curvatures}
\begin{align}
	\cR_{\ha\hb}{}^{\hc\hd} &= -2\d_{[\ha}{}^{\hc}\d_{\hb]}{}^{\hd} \,,
	\\
	\cR_{j}{}^{\bh i }{}_{\ah}{}^{\ha\hb} &= 4\d^{i}{}_{j}(\S^{\ha\hb})_{\ah}{}^{\bh} \,,
	\\
	\cR^{i}{}_{\ah j}{}^{\bh}{}_{k}{}^{l} &= 
	-4\d_{\ah}{}^{\bh}\d^{i}{}_{k}\d^{l}{}_{j}
	\,,
	\\
	(\cR_{\sU(1)})^{i}{}_{\ah j}{}^{\bh} &= \frac{\ri(\N-4)}{\N}\d^{i}{}_{j}\d_{\ah}{}^{\bh}\,. 
\end{align}
\esubeq

\subsection{Covariant derivatives}
In this subsection we will use the results of the previous subsection to detail the construction of covariant derivatives on AdS superspace and determine their algebra.

Associated with each generator $K_{\hA}$ is a vector field $E_{\hA}\,,$ and each generator $H_{\hat{I}}$ a connection $\bm{\F}^{\hat{I}}\,.$ 
To determine $E_{\hA}$ it is useful to introduce the vielbein supermatrix $E_{\hM}{}^{\hA}$ used implicitly above:
\bsubeq
\begin{align}
	\tb{E}^{\hA} &= \bm{\cE}^{\hM} E_{\hM}{}^{\hA} \,,
	\\
	\bm{\cE}^{\hM} &= \tb{E}^{\hA} E_{\hA}{}^{\hM} \,,
\end{align}
\esubeq
with $\bm{\cE}^{\hM}$ given by
\begin{align}
	\bm{\cE}^{\hM} &= (\text{d}x^{\hm}\,, \text{d}\q_{i}{}^{\mh}\,, \text{d}\bar{\q}^{i}{}_{\mh})\,.
\end{align}
We then have 
\begin{align}
	E_{\hA} = E_{\hA}{}^{\hM} \partial_{\hM}\,, \qquad 
	\partial_{\hM} = (\partial_{\hm}\,, \frac{\partial}{\partial \q_{i}{}^{\mh}}\,, \frac{\partial}{\partial \bar{\q}^{i}{}_{\mh}})\,. 
\end{align}

Now we define the covariant derivatives
\begin{align}
	\cD_{\hA} := (\cD_{\ha}\,, \cD_{\ah}{}^{i}\,, \bar{\cD}^{\ah}{}_{i})\,. 
\end{align}
They take the form 
\begin{align} \label{cd form}
	\cD_{\hA} = E_{\hA} + \frac{1}{2}\F_{\hA}{}^{\ha\hb}M_{\ha\hb} + \F_{\hA \, i}{}^{j}\mathbb{J}^{i}{}_{j} + \ri (\F_{\sU(1)}) _{\hA} \mathbb{Y}\,,
\end{align}
and their algebra is given by 
\begin{align}
	[\cD_{\hA}\,, \cD_{\hB}\} = -\cT_{\hA\hB}{}^{\hC}\cD_{\hC} + \frac{1}{2}\cR_{\hA\hB}{}^{\hc\hd}M_{\ha\hb} + \cR_{\hA\hB \, k}{}^{l}\mathbb{J}^{k}{}_{l} + \ri(\cR_{\sU(1)})_{\hA\hB}\mathbb{Y}\,. 
\end{align}
Explicitly, we have
\bsubeq \label{ads cd alg}
\begin{align}
	[\cD_{\ha}\,, \cD_{\hb}] &= 
	-M_{\ha\hb} 
	\,,
	\\
	[\cD_{\ha}\,, \cD_{\ah}{}^{i}] &= 
	\frac{\ri}{2}(\g_{\ha})_{\ah}{}^{\bh}\cD_{\bh}{}^{i} 
	\,,
	\\
	[\cD_{\ha}\,, \bar{\cD}^{\ah}{}_{i}] &= 
	\frac{\ri}{2}(\g_{\ha})^{\ah}{}_{\bh}\bar{\cD}^{\bh}{}_{i}
	\,,
	\\
	\{\cD_{\ah}{}^{i} \,, \bar{\cD}^{\bh}{}_{j} \} &= 
	-2\ri\d^{i}{}_{j}(\g^{\ha})_{\ah}{}^{\bh}\cD_{\ha}
	+ 4\d^{i}{}_{j} M_{\ah}{}^{\bh} 
	\\
	\notag 
	& \quad 
	- 4\d_{\ah}{}^{\bh}\mathbb{J}^{i}{}_{j} - \frac{(\N-4)}{\N}\d^{i}{}_{j}\d_{\ah}{}^{\bh}\mathbb{Y} 
	\,.
\end{align}
\esubeq

In the $\N=1$ case, we can present the above algebra in a different basis, in line with the conventional approach to $5$D superspaces in the literature. 
Let us combine the derivatives $\cD_{\ah}$ and $\bar{\cD}_{\ah}$ into the following $\sSU(2)$ doublet
\begin{align}
	\tilde{\cD}_{\ah}^{\underline{i}}\,, \qquad  \tilde{\cD}_{\ah}^{\underline{1}} = \cD_{\ah}\,, 
	\qquad \tilde{\cD}_{\ah}^{\underline{2}} = \bar{\cD}_{\ah}\,, \qquad \underline{i}\,, \underline{j} = \underline{1}\,, \underline{2}\,.  
\end{align}
The corresponding local coordinates $\q_{\ah}$ and $\bar{\q}_{\ah}$ are combined in a similar way and satisfy a pseudo-Majorana condition.
The $\sSU(2)$ indices are raised and lowered by $\ve^{\underline{ij}}$ and $\ve_{\underline{ij}}$, $\ve^{\underline{12}} = \ve_{\underline{21}} = 1$, as $\tilde{\cD}_{\ah \underline{i}} = \ve_{\underline{ij}}\tilde{\cD}_{\ah}^{\underline{j}}\,.$  
We then have the following relations
\bsubeq
\begin{align}
	[\cD_{\ha}\,, \tilde{\cD}^{\ui}_{\ah}] &= \frac{\ri}{2}(\s_{3})^{\ui}{}_{\uj}(\g_{\ha})_{\ah}{}^{\bh}\tilde{\cD}^{\uj}_{\bh}
	\,,
	\\
	\{\tilde{\cD}_{\ah}^{\underline{i}}\,, \tilde{\cD}_{\bh}^{\underline{j}} \} &= 
	-2\ri \ve^{\underline{ij}} (\g^{\ha})_{\ah \bh}\cD_{\ha} - 4(\s_{3})^{\underline{ij}} M_{\ah\bh} - 3\ve_{\ah\bh}\ve^{\underline{ij}} \mathbb{Y}\,,
\end{align}
\esubeq
and our algebra of covariant derivatives coincides with that of \cite{Kuzenko:2007aj}, see equations (2.21) and (7.49).


\section{AdS superspaces with internal bosonic dimensions} \label{projective}

The analysis of the previous sections made use of \textit{even} supertwistors. 
One can also consider \textit{odd} supertwistors,
\begin{align}
	\bm{\X} = (\bm{\X}_{A}) = \left(\begin{array}{c}
		\bm{\X}_{\ah} \\
		\hline\hline
		\bm{\X}_{i}
	\end{array}\right)\,,
\end{align}
with opposite Grassmann parities 
\begin{align}
	\e(\bm{\X}_{A}) = 1 + \e_{A} \qquad \text{(mod 2)}\,. 
\end{align}
The superspace AdS$^{5|8\N}$ can be extended to AdS$^{5|8\N} \times \mathbb{X}^{\N}_{m}$, where the internal space $\mathbb{X}^{\N}_{m}$ is realised in terms of $m \leq \N$ odd supertwistors. 
In the $\N$-extended case, there are many choices for the space $\mathbb{X}^{\N}_{m}$, corresponding to the various ways in which the $m$ odd supertwistors $\bm{\X}^{\ti}\,, \ti = 1\,,\ldots\,,m$ can be constrained.
In all cases, the odd supertwistors are required to be orthogonal to the quartet $\bm{T}^{\tmu}$,
which means that they contribute no Grassmann odd degrees of freedom.
The remaining constraints on the odd supertwistors should be chosen in such a way that the space 
AdS$^{5|8\N} \times \mathbb{X}^{\N}_{m}$ is a homogeneous space of $\sSU(2,2|\N)$. 
This implies that $\mathbb{X}^{\N}_{m}$ is a homogeneous space of $\sSU(\N)$, and can therefore be realised as a coset space 
\begin{align}
	\mathbb{X}^{\N}_{m} = \frac{\sSU(\N)}{\cG}\,, 
\end{align}
where $\cG$ is a subgroup of $\sSU(\N)$. 
We will outline a few possibilities below. 

Here we consider the superspaces AdS$^{5|8\N} \times \mathbb{X}_{1}^{\N}$. 
Let us accompany the quartet $\bm{T}^{\tmu}$ with a single odd supertwistor, $\bm{\X}$, subject to the orthonormality conditions 
\bsubeq
\begin{align}
	\braket{\bm{T}^{\tmu}}{\bm{\X}} &= 0\,, \label{even odd ip}
	\\
	\braket{\bm{\X}}{\bm{\X}} &= -1\,. \label{odd odd ip}
\end{align}
\esubeq
The second condition implies that the body of $\bm{\X}_{i}$ is non-zero. 
For $\N=1$, the internal space is given by $S^{1}$, whereas for $\N>2$ it can be shown that 
\begin{align} \label{x1n space}
	\mathbb{X}_{1}^{\N} = \frac{\sSU(\N)}{\sSU(\N-1)}\,, \qquad \N>2\,. 
\end{align}
As a consistency check, we can compare the dimension of the space \eqref{x1n space} with the degrees of freedom of the supertwistor $\bm{\X}$. 
The latter has $2\N - 1$ even degrees of freedom, because of the $2\N$ real components and the constraint \eqref{odd odd ip}. 
This coincides with the $\N^{2} - 1 - ((\N-1)^{2}-1) = 2\N-1$ dimensions of \eqref{x1n space}.  

For $\N=2$, we can modify the construction as follows. 
First, we relax the condition \eqref{odd odd ip}. We still require that the body of $\bm{\X}_{i}$ is non-zero. 
Then, we define the supertwistor $\bm{\X}$ modulo the equivalence relation 
\begin{align}
	\bm{\X} \sim d ~ \bm{\X} \,, \qquad d \in \mathbb{C} - \{0\}\,.  
\end{align}
The superspace obtained can be seen to be AdS$^{5|16} \times \mathbb{C}P^{1}$. It is a homogeneous space for $\sSU(2,2|2)\,.$


\subsection{The $\N=4\,, m=4$ case}

As outlined above, for a given $\N$, the maximal number of odd supertwistors we can consider is $\N$. 
With this in mind, for the $\N=4$ case let us introduce the quartet $\bm{\X}^{\ti}\,, ~ \ti = 1\,,2\,,3\,,4\,,$ subject to the orthonormality conditions
\bsubeq
\begin{align}
	\braket{\bm{T}^{\tmu}}{\bm{\X}^{\ti}} &= 0\,,
	\\
	\braket{\bm{\X}^{\ti}}{\bm{\X}^{\tj}} &= -\d^{\ti\tj}\,,
\end{align}
\esubeq
and defined modulo the eqiuvalence relation 
\begin{align} \label{s5 equiv}
	\bm{\X}^{\ti} \sim \bm{\X}^{\tj} \g_{\tj}{}^{\ti}\,, \qquad \g \in \sU(1) \times \sUSp(4)\,. 
\end{align}
Making use of the equivalence relations \eqref{susy u1} and \eqref{s5 equiv}, the $\sU(1)$ freedom on the even and odd supertwistors can be fixed such that the supermatrix  formed by 
$(\bm{T}^{\tmu} \,, \bm{\X}^{\ti})$ has unit Berezinian.
This means that they can be represented as
\begin{align}
	(\bm{T}^{\tmu}\,, \bm{\X}^{\ti}) := \bm{\mathbb{T}} = \left(\begin{array}{c||c}
		\bm{T}_{\ah}{}^{\tmu} & \bm{\X}_{\ah}{}^{\ti} \\
		\hline \hline 
		\bm{T}_{i}{}^{\tmu} & \bm{\X}_{i}{}^{\ti}
	\end{array}\right) 
	\in \sSU(2,2|4)
	\,. 
\end{align}
Furthermore, the equivalence relations that preserve this choice can be written as 
\begin{align} \label{s5 big equiv}
	\bm{\mathbb{T}} \sim \bm{\mathbb{T}} \left(\begin{array}{c||c}
		\re^{\ri\vf} \l & ~0~
		\\
		\hline \hline 
		~0~ & \g
	\end{array}\right)\,,
	\qquad 
	\re^{4\ri\vf} = \det (\g)
	\,. 
\end{align}
It can be shown that the resulting superspace is
\begin{align} \label{ads times s5}
	\frac{\sSU(2,2|4)}{\sUSp(2,2)\times \sU(1) \times\sUSp(4)}\,. 
\end{align}
To prove \eqref{ads times s5}, we note that the simplest  base point is 
\begin{align}
	\bm{\mathbb{T}}^{(0)} &= \left(\begin{array}{c||c}
		\id_{4} & 0 
		\\
		\hline \hline 
		0 & \id_{4}
	\end{array}\right)\,. 
\end{align}
The stabiliser of $\bm{\mathbb{T}}^{(0)}$ consists of those supergroup elements that map $\bm{\mathbb{T}}^{(0)}$ to an equivalent point with respect to \eqref{s5 big equiv}. 
Since the stabiliser of $\bm{\mathbb{T}}^{(0)}$ is a subgroup of $\sSU(2,2|4)$, its elements have unit Berezinian and take the form 
\bsubeq
\begin{align}
	h = \left(\begin{array}{c||c}
		\re^{\ri\vf}\l & ~0~ \\
		\hline \hline 
		~0~ & \g
	\end{array}\right)\,, 
\end{align}
with 
\begin{align}
	\l \in \sUSp(2,2)\,, \quad \g \in \sU(1) \times \sUSp(4)\,, \qquad 
	\re^{4\ri\vf} = \det (\g)\,.  
\end{align}
\esubeq
It is therefore isomorphic to 
\begin{align}
	\sUSp(2,2) \times \sU(1) \times \sUSp(4)\,,
\end{align}
and we arrive at \eqref{ads times s5}. 
This situation describes the superspace AdS$^{5|32} \times \mathbb{X}_{4}^{4}\,,$ where the internal space is given by
\begin{align}
	\mathbb{X}_{4}^{4} = \frac{\sSU(4)}{\sUSp(4)} = \frac{\sSO(6)}{\sSO(5)}
	\,,
\end{align}
where the latter relation may be established using arguments analogous to those given in subsection \ref{bt realisation}. Thus the space $\mathbb{X}_{4}^{4}$ is equivalent to $S^{5}$. 

We can also introduce bi-supertwistors for the space $\mathbb{X}^{4}_{4}$. 
Let us introduce the following supermatrix 
\begin{align}
	\bm{\z}_{AB} = r\bm{\X}_{A}{}^{\ti}\bm{\X}_{B}{}^{\tj}C_{\ti\tj}\,,
\end{align}
where $C_{\ti\tj}$ is given by 
\begin{align}
	(C_{\ti\tj}) = \left(\begin{array}{cc}
		\ve & ~0~
		\\
		~0~ & \ve
	\end{array}\right)\,, 
\end{align} 
and $r$ is the $S^{5}$ radius.
The supermatrix $\bm{\z}_{AB}$ is defined modulo the equivalence relation
\begin{align}
	\bm{\z}_{AB} \sim \re^{\ri\vf}\bm{\z}_{AB}\,, \qquad \vf \in \mathbb{R}\,,
\end{align}
which is the remnant of the  
equivalence transformations  \eqref{s5 equiv}. 
Because the quartet $\bm{\X}^{\ti}$ has opposite Grassmann parity to $\bm{T}^{\tmu}$, the supermatrix $\bm{\z}_{AB}$ has the following symmetry property
\begin{align}
	\bm{\z}_{AB} = -(-1)^{(1+\e_A)(1+\e_B)}\bm{\z}_{BA}\,.
\end{align}
It also satisfies the conditions
\bsubeq
\begin{align}
	\bar{\bm{X}}{}^{AC}\bm{\z}_{CB} &= 0\,,
	\\
	\bar{\bm{\z}}{}^{AB}\bm{\z}_{BA} &= 4r^{2}\,,
\end{align}
\esubeq
where $\bar{\bm{\z}}{}^{AB}$ is defined in the same way as \eqref{dual sbt def}. 

The bi-supertwistors $\bm{X}_{AB}$ and $\bm{\z}_{AB}$ can be combined in the following way. 
Let us introduce the supermatrix 
\begin{align}
	\cX = \left(\begin{array}{c||c}
		\mathfrak{C}_{\tmu\tnu} & 0
		\\
		\hline\hline
		0 & C_{\ti\tj}
	\end{array}\right) \in \sSU(2,2|4)\,.
\end{align}
Then, we find the following 
\begin{align}
	\U &:= \bm{\mathbb{T}} \cX \bm{\mathbb{T}}^{\T} = \ell^{-1}\bm{X} + r^{-1}\bm{\z}\,, \qquad 
	\bm{X} = (\bm{X}_{AB})\, , \quad \bm{\z} = (\bm{\z}_{AB})\,.
\end{align}
Due to the $\sU(1)$ freedom enjoyed by $\bm{X}$ and $\bm{\z}$, the supermatrix $\U$ is defined modulo 
\begin{align}
	\U \sim \re^{\ri\vf}\U\,, \qquad \vf \in \mathbb{R}\,. 
\end{align}
Furthermore, it satisfies the following condition
\begin{align}
	\U^{\dag}\bm{\O}\U = \bm{\O}\,. 
\end{align}
%


\section{Conclusion} \label{conclusion}

In this paper we presented the supersymmetric generalisation of the embedding formalism for AdS$_5$. 
Specifically, we proposed supertwistor and bi-supertwistor realisations for AdS$^{5|8\N}$.
Making use of the coset construction, we studied the superspace geometry of AdS$^{5|8\N}$, and showed that our results are consistent with the known case for the $\N=1$ superspace AdS$^{5|8}$. 
For AdS$^{5|8\N}$, we introduced  Poincar\'e-like superspace coordinates and computed the bi-supertwistor in this coordinate system. 
We also extended the (bi-)supertwistor realisations to superspaces of the form AdS$^{5|8\N} \times \mathbb{X}^{\N}_{m}$, where $\mathbb{X}^{\N}_{m}$ is
a homogeneous space of $\sSU(\N)$. 
In particular, for the $\N=m=4$ case, we derived the necessary constraints for the (bi-)supertwistors to describe the AdS$_5 \times S^{5}$ superspace \eqref{ads times s5}.
It would be interesting to recast the type IIB superstring action in AdS$_5 \times S^{5}$ background \cite{Metsaev:1998it} in terms of the bi-supertwistors $\bm{X}_{AB}$ and $\bm{\x}_{AB}$. 
This will be discussed elsewhere.

One of the main results of our paper is the new model \eqref{superparticle model} for a superparticle propagating in AdS$^{5|8\N}$. The distinguished feature of this model is that it involves two independent two-derivative terms. 
It is of interest to see how these two-derivative structures may originate within a supergravity setting. Such a setting is available in the $\cN=1$ case.
We will now turn to re-deriving the superparticle model from the perspective of the $\cN=1$ supergravity framework in five dimensions as reviewed in \cite{Kuzenko:2008kw}.

Let $z^{\hM} = (x^{\hm}\,, \q_{\ui}^{\mh})$, with $\ui = \underline{1}\,, \underline{2}$ and $\overline{\q_{\ui}^{\mh}} = \q^{\ui}_{\mh}$, be local coordinates on the $\N=1$ superspace AdS$^{5|8}\,.$
The  covariant derivatives of AdS$^{5|8}$, $\cD_A = \big( \cD_{\ha} \, ,\cD_{\ah}^{\ui}\big)$, obey the graded commutation relations 
\bsubeq
\begin{align}
	\{\cD_{\ah}^{\ui}\,,\cD_{\bh}^{\uj}\} &= -2\ri\ve^{\underline{ij}}\cD_{\ah\bh} + 4\ri\cS^{\underline{ij}}M_{\ah\bh} 
	+ 3\ri\ve_{\ah\bh}\ve^{\underline{ij}}\cS^{\uk\ul}J_{\uk\ul}\,,
	\\
	[\cD_{\ha}\,, \cD_{\ah}^{\ui}] &= \frac{1}{2}(\g_{\ha})_{\ah}{}^{\bh}\cS^{\ui}{}_{\uj}\cD_{\bh}^{\uj}\,.
\end{align}
\esubeq
Here $M_{\ah\bh} = M_{\bh\ah} $ is the Lorentz generator, $J_{\uk\ul}\ = J_{\ul\uk}$ the $\sSU(2)$ generator,
and $\cS^{\underline{ij}}$ is a non-vanishing 
covariantly constant 
tensor superfield with the algebraic properties:
\begin{align} \label{s prop}
	\cS^{\underline{ij}} = \cS^{\underline{ji}}\,, \qquad \overline{\cS^{\underline{ij}}} = \cS_{\underline{ij}}
	= \ve_{\underline{ik}} \ve_{\underline{jl}}\cS^{\underline{kl}} 
	 ~\implies ~ \cS^{\underline{ik}}\cS_{\underline{kj}} = \cS^{2}\d^{\ui}{}_{\uj}\,, \quad \cS^{2} = \frac{1}{2}\cS^{\underline{ij}}\cS_{\underline{ij}}\,. 
\end{align}
In matrix notation, the properties of $\mathbb{S} = \cS^{-1}(\cS^{\underline{ij}})$ can be written as 
\begin{align} \label{MatrixConditions}
	\mathbb{S} = \mathbb{S}^{\T}\,, \qquad \mathbb{S}^{\dag}\mathbb{S} = \id_{2}\,, 
	\qquad \mathbb{S}^{\dag} = \ve \mathbb{S}\ve^{-1} \,.
\end{align}
It then follows that $\mathbb{S} $ admits the representation\footnote{Had we omitted the reality condition in  
\eqref{MatrixConditions},
$\mathbb{S}^{\dag} = \ve \mathbb{S}\ve^{-1} $,  the general solution would be $\mathbb{S} = U U^{\T}$ with  $U \in \sU(2)$, 
 see, e.g., \cite{Zumino:1962smg}. }
\begin{align}
	\mathbb{S} = U U^{\T}\,, \qquad U \in \sSU(2)\,.
\end{align} 
Its meaning is very simple. Specifically, the space of solutions to \eqref{MatrixConditions} can be identified with a unit two-sphere, $S^2$, which is a homogeneous space for $\sSU(2$). In particular, applying an $\sSU(2)$ transformation allows one to choose $\cS^{\underline{ij}} = \d^{\underline{ij}}$. There is  a different choice for  $\cS^{\underline{ij}} $ allowing one to make direct correspondence with  \cite{Kuzenko:2008kw}.
Making use of a unitary transformation, $\cS^{\underline{ij}}$ can be brought to the form 
\begin{align} \label{sigma 3}
	\cS^{\underline{ij}} = \ri \cS(\s^{3})^{\underline{ij}}\,,
\end{align}
and the resulting algebra of covariant derivatives coincides with \eqref{ads cd alg} provided one fixes $\cS = 1$. 
As outlined in \cite{KKR2}, the cost of making such a choice is the loss of a conformally flat frame.
Unless otherwise stated, we will not impose the gauge condition \eqref{sigma 3} below. 

Making use of $\cS^{\underline{ij}}$, we can introduce a one-parameter deformation of the supersymmetric interval $\eta_{\ha\hb}E^{\ha}E^{\hb}$ 
\begin{align}
	\rd s^{2} = \eta_{\ha\hb}E^{\ha}E^{\hb} + \frac{\ri\o}{\cS^{2}}\cS^{\underline{ij}}\ve_{\ah\bh}E_{\ui}^{\ah}E_{\uj}^{\bh}\,,
\end{align}
for a real dimensionless parameter $\o\,.$\footnote{The four-dimensional analogue of this deformation was considered recently in \cite{KKR2} for arbitrary $\N$.}
This can be written as 
\begin{align}
	\rd s^{2} = E^{\hA}\eta_{\hA\hB}E^{\hB}\,,
\end{align}
where the supermatrix $\eta_{\hA\hB}$ is defined as 
\begin{align}
	\eta_{\hA\hB} = \left(
	\begin{array}{c||c}
		\eta_{\ha\hb} & 0 \\
		\hline \hline
		~0~ & \frac{\ri \o}{\cS^{2}}\cS^{\underline{ij}}\ve_{\ah\bh}
	\end{array}
	\right)\,.
\end{align}
Given such a deformed interval, it follows that there is a superparticle model 
\begin{align} \label{def model}
	S = \frac{1}{2}\int \rd \t \frak{e}^{-1} \left\{ \dt{E}{}^{\hA}\eta_{\hA\hB}\dt{E}{}^{\hB} - (\frak{e}m)^{2}\right\}\,, \qquad \dt{E}{}^{\hA} = \frac{\rd z^{\hM}}{\rd\t} E_{\hM}{}^{\hA}\,,
\end{align}
where $\frak{e}$ is the einbein and $m$ is the mass. 
For the case $\o = 0$ we recover the standard superparticle model in five dimensions.

In order to make contact with the model \eqref{superparticle model}, one must make use of the results of \cite{Kuzenko:2007aj}, in which the components of the supervielbein for the AdS$^{5|8}$ coset superspace are provided in the Poincar\'e-like coordinate system discussed in section \ref{section 4.2}. 
Then, in the gauge \eqref{sigma 3}, the deformation takes the form
\begin{align}
	\frac{\ri\o}{2\cS^{2}}\cS^{\underline{ij}}\ve_{\ah\bh}\dt{E}_{\ui}{}^{\ah}\dt{E}_{\uj}{}^{\bh} &= 
	\frac{\o}{\cS^{2}}\bigg(
	\dot{\q}^{\a}\dot{\eta}_{\a} + \dot{\bar{\q}}_{\ad}\dot{\bar{\eta}}^{\ad} + 2\dot{\q}^{2}\eta^{2} + 2\dot{\bar{\q}}^{2}\bar{\eta}^{2} + 4\dot{\q}^{\a}\eta_{\a}\bar{\eta}_{\ad}\dot{\bar{\q}}^{\ad} 
	- 2\eta^{2}\bar{\eta}^{2}\dt{\P}{}^{2}
	\\
	\notag & \qquad \qquad  + 4\ri\dt{\P}{}^{\ad\a}(\bar{\eta}^{2}\eta_{\a}\dot{\bar{\q}}_{\ad} - \eta^{2}\dot{\q}_{\a}\bar{\eta}_{\ad}) + \ri\dt{\P}{}^{\ad\a}(\bar{\eta}_{\ad}\dot{\eta}_{\a} - \dot{\bar{\eta}}_{\ad}\eta_{\a}) 
	 \bigg)\,, 
\end{align}
where $\dt{\P}{}^{a} = \dot{x}^{a} + \ri(\q\s^{a}\dot{\bar{\q}} - \dot{\q}\s^{a}\bar{\q})$.
This can be shown to coincide with the $\a$-term in \eqref{superparticle model} provided one fixes 
\begin{align}
	\a = \frac{\o}{4\cS^{2}}\,.
\end{align}

An important feature of massless models of this kind is the presence of the so-called $\k$-symmetry, which was introduced for the superparticle in \cite{Siegel:1983hh} and generalised to the superstring in \cite{Green:1983wt}, see, e.g., \cite{Sezgin:1993xg} for a review.  
The $\k$-symmetry is most conveniently described in a conformally flat frame, which was derived for AdS$^{5|8}$ in \cite{Kuzenko:2008kw}. 
In such a frame, the components of the supervielbein are
\bsubeq
\begin{align}
	E^{\ha} &= \re^{-2\s}e^{\ha}\,, \qquad e^{\ha} = \rd x^{\ha} + \ri \rd \q_{\ui}^{\ah}(\g^{\ha})_{\ah\bh}\q^{\bh\ui}\,,
	\\
	E_{\ui}^{\ah} &= \re^{-\s}\left\{\rd \q_{\ui}^{\ah} + \ri(D_{\bh \ui}\s)e^{\bh\ah} \right\}\,, 
\end{align}
\esubeq
where $e^{\ha}$ is the flat supersymmetric one form in five dimensions, $\s$ is the super-Weyl parameter, and $D_{\ah}^{\ui} = \partial_{\ah}^{\ui} - \ri(\g^{\ha})_{\ah\bh}\q^{\bh \ui}\partial_{\ha}$ is the flat spinor covariant derivative.
Then, for $m = \o = 0$, the model \eqref{def model} takes the form 
\begin{align}
	S = \frac{1}{2}\int \rd\t\frak{e}^{-1}\re^{-4\s} \eta_{\ha\hb} \dt{e}{}^{\ha}\dt{e}{}^{\hb}\,, \qquad \dt{e}{}^{\ha} = \dot{x}{}^{\ha} + \ri\dot{\q}_{\ui}^{\ah}(\g^{\ha})_{\ah\bh}\q^{\bh\ui}\,. 
\end{align} 
and is invariant under the following local transformations 
\bsubeq
\begin{align}
	\d\q_{\ui}^{\ah} &= \ri\dt{e}{}^{\ah\bh}\k_{\bh\ui}\,,
	\\
	\d x^{\ha} &= \ri \q_{\ui}^{\ah}(\g^{\ha})_{\ah\bh}\d\q^{\bh\ui}\,,
	\\
	\d\frak{e} &= -4\frak{e}\left\{ \dot{\q}_{\ui}^{\ah}\k_{\ah}^{\ui} + \d\q_{\ui}^{\ah}D_{\ah}^{\ui}\s  \right\}\,, 
\end{align}
\esubeq
where $\k_{\ah}^{\ui} = \k_{\ah}^{\ui}(\t)$ is Grassmann-odd. 
It is an interesting problem to extend these considerations beyond $\N=1$ and recast the local $\k$-symmetry in terms of supertwistors. 
This will be discussed elsewhere.

\noindent
{\bf Acknowledgements:}\\
We are grateful to Emmanouil Raptakis for useful discussions. 
 The work of SMK  is supported in part by the Australian Research Council, project DP230101629.
The work of NEK is supported by the Australian Government Research Training Program Scholarship.

\appendix

\section{Spinors in diverse dimensions} \label{spin}

In this appendix we collect those results of the spinor formalisms in $3+1$, $4+2$ and $4+1$ dimensions, which are used in our work. 

\subsection{Spinors in $3+1$ dimensions} \label{two components}

Our two-component spinor conventions coincide with those in the books \cite{WB, Buchbinder:1998qv}, 
to which the reader is referred to for a complete discussion. 

Consider the space $\mathbb{R}^{3,1}$ equipped with metric
\begin{align}
	\eta_{ab} = \text{diag}(-1\,,+1\,,+1\,,+1)\,, \qquad \qquad a = 0,1,2,3\,. 
\end{align}
The gamma matrices in $3+1$ dimensions satisfy 
\begin{align}
	\{\g_{a}\,,\g_{b}\} = -2\eta_{ab}\id_{4}\,,
\end{align}
and can be chosen as
\begin{align} \label{4d gm}
	\g_{a} = \left(\begin{array}{cc}
		0 & \s_{a} \\
		\tilde{\s}_{a} & 0
	\end{array}\right)\,, 
\end{align}
with 
\begin{align}
	\s_{a} = (\id_{2}\,, \vec{\s}) \equiv (\s_{a})_{\a\ad}\,, \qquad \tilde{\s}_{a} = (\id_{2}\,, -\vec{\s}) \equiv (\tilde{\s}_{a})^{\ad\a}\,. 
\end{align}
The matrix $\g_{5} = -\ri\g_0\g_1\g_2\g_3$ then takes the form 
\begin{align}
	\g_{5} = \left(\begin{array}{cc}
		\id_{2} & 0 \\
		0 & -\id_{2}
	\end{array}\right)\,.
\end{align}
The Lorentz generators in the spinor representation are 
\begin{align}
	\S_{ab} = -\frac{1}{4}[\g_{a}\,,\g_{b}]\,.
\end{align}
They are given by 
\bsubeq 
\begin{align}
	(\S_{ab})_{\ah}{}^{\bh} &= \left(\begin{array}{cc}
		(\s_{ab})_{\a}{}^{\b} & 0
		\\
		0 & (\tilde{\s}_{ab})^{\ad}{}_{\bd}
	\end{array}\right)\,,
	\\
	(\s_{ab})_{\a}{}^{\b} &= -\frac{1}{4}(\s_{a}\tilde{\s}_{b}-\s_{b}\tilde{\s}_{a})_{\a}{}^{\b} \,,
	\\
	(\tilde{\s}_{ab})^{\ad}{}_{\bd} &= -\frac{1}{4}(\tilde{\s}_{a}\s_{b}-\tilde{\s}_{b}\s_{a})^{\ad}{}_{\bd}\,.
\end{align}
\esubeq
Two-component spinor indices are raised and lowered using the spinor metrics
\begin{subequations} \label{epsilon def}
\bea
\ve^{\ab} &=& - \ve^{\b \a} ~, \qquad \ve_{\ab} = - \ve_{\b \a} ~, \qquad \ve^{12}= \ve_{21} =1~; \\
\ve^{\ad \bd} &=& - \ve^{\bd \ad} ~, \qquad \ve_{\ad \bd} = - \ve_{\bd \ad} ~, \qquad \ve^{\dot 1 \dot 2}= \ve_{\dot 2 \dot 1} =1~,
\eea
\end{subequations}
by the rules: 
\begin{align} \label{raise and lower}
	\psi^{\a} := \ve^{\ab}\psi_{\b}\,, \qquad \psi_{\a} = \ve_{\ab}\psi^{\b}\,; \qquad 
	\bar \phi^{\ad} := \ve^{\ad \bd}\bar \phi_{\bd}\,, \qquad \bar \phi_{\ad} = \ve_{\ad \bd} \bar \phi^{\bd}\,.
\end{align}

The charge conjugation matrix, $\mathcal{C} $, is defined by 
\bsubeq \label{4d charge conjugation}
\begin{align} 
	\mathcal{C}^{-1}\g_{a}\mathcal{C} = -\g_{a}^{\T}
	\quad \implies \quad 
	\mathcal{C}^{-1}\g_{5}\mathcal{C} = \g_{5}^{\T}\,.
\end{align}
It has the properties
\begin{align}
	\mathcal{C}^{\dag} = \mathcal{C}^{\T} = -\mathcal{C} = \mathcal{C}^{-1}\,,
\end{align}
and can be chosen as 
\begin{align}
	\mathcal{C} = \left(
	\begin{array}{cc}
		\ve_{\ab} & 0 \\
		0 & \ve^{\ad\bd}
	\end{array}
	\right)\,. 
\end{align}
\esubeq

A number of useful properties are satisfied by the matrices $\s_{a}$ and $\tilde{\s}_{a}$. 
They are: 
\bsubeq 
\begin{align}
	(\s_a \tilde{\s}_{b} + \s_{b}\tilde{\s}_{a})_{\a}{}^{\b} &= -2\eta_{ab}\d_{\a}{}^{\b}\,,
	\\
	(\tilde{\s}_{a}\s_{b} + \tilde{\s}_{b}\s_{a})^{\ad}{}_{\bd} &= -2\eta_{ab}\d^{\ad}{}_{\bd}\,,
	\\
	\text{Tr}(\s_{a}\tilde{\s}_{b}) &= -2\eta_{ab}\,,
	\\
	(\s^{a})_{\a\ad}(\tilde{\s}_{a})^{\bd\b} &= -2\d_{\a}{}^{\b}\d_{\ad}{}^{\bd}\,. 
\end{align}
\esubeq

Associated with a four-vector $x^{a} \in \mathbb{R}^{3,1}$ are the Hermitian matrices 
\begin{align}
	x_{\a\ad} = x^{a}(\s_{a})_{\a\ad}\,, \quad x^{\ad\a} = x^{a}(\tilde{\s}_{a})^{\ad\a}\,. 
\end{align}
It is sometimes convenient to use the condensed notation 
\begin{align}
	x := (x_{\a\ad})\,, \qquad \tilde{x}:= (x^{\ad\a})\,. 
\end{align}

\subsection{Spinors in $4+2$ dimensions} \label{spinor conventions}

In this appendix we will detail the spinor conventions for $\sSO_0(4,2)\,.$
Consider the space $\mathbb{R}^{4,2}$ parametrised by coordinates $x^{\ua}\,, ~ \ua = 0,1,2,3,4,5\,,$ and with metric 
\begin{align}
	\eta_{\ua\ub} = \text{diag}(-1,+1,+1,+1,+1,-1)\,. 
\end{align}
The corresponding gamma matrices, $\G_{\ua}$, obey the anti-commutation relations
\begin{align} \label{6D clifford}
	\{\G_{\ua},\G_{\ub}\} = -2\eta_{\ua\ub}\id_{8}\,.
\end{align}
We can choose $\G_{\ua}$ in the so-called Weyl repersentation
\begin{align}
	\G_{\ua} = \left(\begin{array}{cc}
		0 & \underline{\S}_{\ua} \\
		\underline{\tilde{\S}}_{\ua} & 0
	\end{array}\right)\,,
\end{align}
with 
\bsubeq
\begin{align}
	\underline{\S}_{\ua} &= \left(\ri\g_{a}\,, \g_{5}, \id_{4}\right) \equiv (\underline{\S}_{\ua})_{\ah\underline{\bh}}\,, 
	\\
	\underline{\tilde{\S}}_{\ua} &= \left(-\ri\g_{a}\,, -\g_{5}\,, \id_{4}\right) \equiv (\underline{\tilde{\S}}_{\ua})^{\underline{\ah}\bh}\,.
\end{align}
\esubeq
Here $\g_{a}$ and $\g_{5}$ are the gamma matrices in $3+1$ dimensions, see appendix \ref{two components}. 

In accordance with \eqref{6D clifford}, the matrices $\underline{\S}$ and $\underline{\tilde{\S}}$ obey the following relations 
\begin{align} \label{reduced 6d clifford}
	\underline{\S}_{\ua}\tilde{\underline{\S}}_{\ub} + \underline{\S}_{\ub}\tilde{\underline{\S}}_{\ua} = -2\eta_{\ua\ub}\id_{4}\,,
	\qquad 
	\tilde{\underline{\S}}_{\ua}\underline{\S}_{\ub} + \tilde{\underline{\S}}_{\ub}\underline{\S}_{\ua} = -2\eta_{\ua\ub}\id_{4}\,. 
\end{align}
The Hermitian conjugation properties of $\G_{\ua}$ are 
\begin{align} \label{big gam conj}
	\G_{\ua}^{\dag} = \G_0\G_5\G_{\ua}\G_0\G_5\,,
\end{align}
which implies the following Hermitian conjugation properties for the matrices $\underline{\S}$ and $\underline{\tilde{\S}}$
\begin{align} \label{u sigma conj}
	(\underline{\S}_{\ua})^{\dag} = \g_{0}\underline{\tilde{\S}}_{\ua}\g_{0}\,, \qquad (\underline{\tilde{\S}}_{\ua})^{\dag} = \g_{0}\underline{\S}_{\ua}\g_{0}\,. 
\end{align}

The Dirac spinor representation is generated by 
\begin{align} \label{Lorentz gens}
	\mathfrak{J}_{\ua\ub} = -\frac{1}{4}[\G_{\ua},\G_{\ub}] = \left(\begin{array}{cc}
		\underline{\S}_{\ua\ub} & 0 \\
		0 & \tilde{\underline{\S}}_{\ua\ub}
	\end{array}\right)\,.
\end{align}
Here we have defined 
\bsubeq \label{underlined lorentz gen}
\begin{align}
	\underline{\S}_{\ua\ub} &:= -\frac{1}{4}\left(\underline{\S}_{\ua}\tilde{\underline{\S}}_{\ub} - \underline{\S}_{\ub}\tilde{\underline{\S}}_{\ua}\right) \equiv (\underline{\S}_{\ua\ub})_{\ah}{}^{\bh} \,, \label{underlined twistor gen}
	\\
	\underline{\tilde{\S}}_{\ua\ub} & := 
	-\frac{1}{4}\left(\tilde{\underline{\S}}_{\ua}\underline{\S}_{\ub} - \tilde{\underline{\S}}_{\ub}\underline{\S}_{\ua}\right) \equiv (\tilde{\underline{\S}}_{\ua\ub})^{\underline{\ah}}{}_{\underline{\bh}}\,. \label{underlined dual gen}
\end{align}
\esubeq
Given \eqref{big gam conj}, the Hermitian conjugation properties of $\mathfrak{J}_{\ua\ub}$ are
\begin{align}
	(\mathfrak{J}_{\ua\ub})^{\dag} = \G_0\G_5(\mathfrak{J}_{\ua\ub})\G_0\G_5\,.
\end{align}
For the matrices $\underline{\S}_{\ua\ub}$ and $\tilde{\underline{\S}}_{\ua\ub}$ we have the following
\begin{align}
	(\underline{\S}_{\ua\ub})^{\dag} = -\g_0(\underline{\S}_{\ua\ub})\g_{0}\,, \qquad 
	(\tilde{\underline{\S}}_{\ua\ub})^{\dag} = -\g_0(\tilde{\underline{\S}}_{\ua\ub})\g_{0}\,. 
\end{align}

A Dirac spinor looks like 
\begin{align} \label{diracspinor}
	\Psi = \left(\begin{array}{c}
		\psi \\
		\f
	\end{array} \right)\,, \qquad \psi = (\psi_{\ah})\,, \quad \f = (\f^{\underline{\ah}})\,. 
\end{align}
Its conjugate is defined as 
\begin{align}
	\bar{\Psi} := -\ri\Psi^{\dag}\G_{0}\G_{5} = 
	\left(\begin{array}{cc}
		\psi^{\dag}\g_{0} \,, & - \f^\dag\g_0 
	\end{array}\right) 
	\,, 
		\quad \psi^{\dag}\g_0 \equiv (\bar{\psi}^{\ah})\,, 
			\quad \phi^\dag\g_0 \equiv (\bar{\phi}_{\underline{\ah}})\,.
\end{align}
An infinitesimal $\sSO_0(4,2)$ transformation acts on $\Psi$ and $\bar{\Psi}$ as 
\bsubeq \label{infinitesimal trf}
\begin{align}
	\d \Psi &= \frac{1}{2}\o^{\ua\ub}\mathfrak{J}_{\ua\ub}\Psi\,, \label{infini 1}
	\\
	\d \bar{\Psi} &= -\frac{1}{2}\bar{\Psi}\o^{\ua\ub}\mathfrak{J}_{\ua\ub}\,. \label{infini 2}
\end{align}
\esubeq
This representation is the sum of two irreducible representations, as is clear from \eqref{Lorentz gens}. The twistor (left Weyl spinor) representation is associated with spinors of the form
\begin{align}
	\Psi_{\text{L}} = \left(\begin{array}{c}
		\psi \\
		0
	\end{array}\right)\,, \qquad \psi = (\psi_{\ah})\,,
\end{align}
such that
\begin{align}
	\G_{7}\Psi_{\text{L}} = \Psi_{\text{L}}\,,
\end{align}
where $\G_{7}$ is defined as 
\begin{align}
	\G_{7} := -\ri\G_0\G_1\G_2\G_3\G_4\G_5 \equiv \left(\begin{array}{cc}
		\id_{4} & 0\\
		0 & -\id_{4}
	\end{array}\right)\,. 
\end{align}
The Dirac conjugate of $\Psi_{\text{L}}$ is given by
\begin{align}
	\bar{\Psi}_{\text{L}} = \left(\begin{array}{cc}
		\bar{\psi} \,, & 0
	\end{array}\right)\,, \qquad \bar{\psi} := \psi^\dag\g_{0} = (\bar{\psi}^{\ah})\,,
\end{align}
and transforms according to the dual twistor representation. Following \eqref{infinitesimal trf}, the infinitesimal $\sSO(4,2)$ transformation laws of $\Psi_{\text{L}}$ and $\bar{\Psi}_{\text{L}}$ are given by
\bsubeq
\begin{align}
	\d \psi_{\ah} &= \frac{1}{2}\o^{\ua\ub}(\underline{\S}_{\ua\ub})_{\ah}{}^{\bh}\psi_{\bh}\,,
	\\
	\d \bar{\psi}^{\ah} &= -\frac{1}{2}\bar{\psi}^{\bh}\o^{\ua\ub}(\underline{\S}_{\ua\ub})_{\bh}{}^{\ah}\,.
\end{align}
\esubeq
In the main body of this paper, we use the matrices $\S_{\ua}$ and $\tilde{\S}_{\ua}$. These are related to the matrices $\underline{\S}_{\ua}$ and $\underline{\tilde{\S}}_{\ua}$ as described below. We introduce the $8\times8$ charge conjugation matrix, $\mathscr{C}$, satisfying
\begin{align}
	\mathscr{C}^{-1}\G_{\ua}\mathscr{C} &= -\G_{\ua}^{\T}\,.
\end{align}
It can be chosen as
\begin{align}
	\mathscr{C} = \left(\begin{array}{cc}
		0 & \g_{5}\mathcal{C} \\
		-\g_{5}\mathcal{C} & 0
	\end{array}\right) &\equiv \left(\begin{array}{cc}
		0 & \mathscr{C}_{\ah}{}^{\underline{\bh}} \\
		\mathscr{C}^{\underline{\ah}}{}_{\bh} & 0
	\end{array}\right)\,,
\end{align}
with $\mathcal{C}$ as the $3+1$ dimensional charge conjugation matrix, \eqref{4d charge conjugation}. 
The inverse of $\mathscr{C}$ is 
\begin{align}
	\mathscr{C}^{-1} = 
	\left(
	\begin{array}{cc}
		0 & -\mathcal{C}^{-1}\g_{5} \\
		\mathcal{C}^{-1}\g_{5} & 0 
	\end{array}\right)
	\equiv 
	\left(\begin{array}{cc}
		0 & (\mathscr{C}^{-1})^{\ah}{}_{\underline{\bh}} \\
		(\mathscr{C}^{-1})_{\underline{\ah}}{}^{\bh} & 0
	\end{array}\right) \,.
\end{align}
We also have
\begin{align} \label{conjugate lorentz}
	\mathscr{C}^{-1}\mathfrak{J}_{\ua\ub}\mathscr{C} = -\mathfrak{J}_{\ua\ub}^{\T}\,. 
\end{align}
Using $\mathscr{C}$, we can define the charge conjugate spinor 
\begin{align}
	\Psi_{\mathscr{C}} := \mathscr{C}\bar{\Psi}^{\T}\,. 
\end{align}
Its infinitesimal transformation rule is
\begin{align} \label{conj trf}
	\d \Psi_{\mathscr{C}} = \frac{1}{2}\o^{\ua\ub}\mathfrak{J}_{\ua\ub}\Psi_{\mathscr{C}}\,,
\end{align}
which coincides with \eqref{infini 1}. 
Consider the contents of the charge conjugate spinor
\begin{align}
	\Psi_{\mathscr{C}} &= \mathscr{C}\bar{\Psi}^{\T} 
	\\
	\notag 
	&= \left(\begin{array}{cc}
		0 & \mathscr{C}_{\ah}{}^{\underline{\bh}} \\
		\mathscr{C}^{\underline{\ah}}{}_{\bh} & 0
	\end{array}\right)
	\left(\begin{array}{c}
		\bar{\psi}^{\bh} \\
		-\bar{\f}_{\underline{\bh}}
	\end{array}\right) 
	\\
	\notag 
	&= \left(\begin{array}{c}
		-\mathscr{C}_{\ah}{}^{\underline{\bh}}\bar{\f}_{\underline{\bh}} \\
		\mathscr{C}^{\underline{\ah}}{}_{\bh}\bar{\psi}^{\bh}
	\end{array}\right)\,.
\end{align}
Since $\Psi_{\mathscr{C}}$ transforms as \eqref{conj trf}, $\bar{\f}_{\ah} := \mathscr{C}_{\ah}{}^{\underline{\bh}}\bar{\f}_{\underline{\bh}} = \bar{\f}_{\underline{\bh}}\mathscr{C}^{\underline{\bh}}{}_{\ah}$ transforms like a twistor. 
This means we can convert underlined indices into not-underlined indices. 

We have so far considered the twistor representation. 
The other (right Weyl spinor) representation is equivalent to the dual twistor representation, as 
\begin{align}
	\f^{\ah} := (\mathscr{C}^{-1})^{\ah}{}_{\underline{\bh}}\f^{\underline{\bh}} = \f^{\underline{\bh}}(\mathscr{C}^{-1})_{\underline{\bh}}{}^{\ah}
\end{align} 
transforms like a dual twistor. This also follows from \eqref{conjugate lorentz}, which yields 
\begin{align}
	-(\underline{\S}^{\T}_{\ua\ub})^{\ah}{}_{\bh} 
	=
	 (\mathscr{C}^{-1})^{\ah}{}_{\underline{\gh}}(\tilde{\underline{\S}}_{\ua\ub})^{\underline{\gh}}{}_{\underline{\dhat}}\mathscr{C}^{\underline{\dhat}}{}_{\bh}\,. 
\end{align}

Now, the matrices $\S_{\ua}$ and $\tilde{\S}_{\ua}$ (with twistor indices) are defined as 
\begin{align} \label{new sigma def}
	(\S_{\ua})_{\ah\bh} := (\underline{\S}_{\ua})_{\ah\underline{\gh}}\mathscr{C}^{\underline{\gh}}{}_{\bh}\,, 
	\qquad (\tilde{\S}_{\ua})^{\ah\bh} := (\mathscr{C}^{-1})^{\ah}{}_{\underline{\gh}}(\underline{\tilde{\S}}_{\ua})^{\underline{\gh}\bh}\,.  
\end{align}
They are antisymmetric
\begin{align}
	(\S_{\ua})_{\ah\bh} = -(\S_{\ua})_{\bh\ah}\,, \qquad (\tilde{\S}_{\ua})^{\ah\bh} = -(\tilde{\S}_{\ua})^{\bh\ah}\,,
\end{align}
and satisfy the relations 
\begin{align} \label{s stilde clifford}
		\S_{\ua}\tilde{\S}_{\ub} + \S_{\ub}\tilde{\S}_{\ua} = -2\eta_{\ua\ub}\id_{4}\,,
	\qquad 
	\tilde{\S}_{\ua}\S_{\ub} + \tilde{\S}_{\ub}\S_{\ua} = -2\eta_{\ua\ub}\id_{4}\,. 
\end{align}
It is useful here to give the explicit form of these matrices:
\begin{align} \label{bt basis def}
	\begin{split}
		\S_{a} &= \left(
		\begin{array}{cc}
			0 & \ri\s_{a}\ve \\
			-\ri \tilde{\s}_{a}\ve^{-1} & ~0~
		\end{array}
		\right)\,,
		\\
		\S_{4} &= \left(
		\begin{array}{cc}
			-\ve^{-1} & ~0~ \\
			0 & -\ve
		\end{array}
		\right)\,,
		\\
		\S_{5} &= \left(
		\begin{array}{cc}
			-\ve^{-1} & ~0~ 
			\\
			0 & \ve
		\end{array}
		\right)\,,
	\end{split}
\quad 
\begin{split}
	\tilde{\S}_{a} &= \left(
	\begin{array}{cc}
		0 & \ri\ve\s_{a} \\
		-\ri\ve^{-1}\tilde{\s}_{a} & ~0~
	\end{array}\right)\,,
\\
\tilde{\S}_{4} &= \left(
\begin{array}{cc}
	\ve & ~0~ \\
	~0~ & \ve^{-1}
\end{array}
\right)\,,
\\
\tilde{\S}_{5} &= \left(
\begin{array}{cc}
	-\ve & ~0~ \\
	~0~ & \ve^{-1}
\end{array}
\right)\,. 
\end{split}
\end{align}
Since the Lorentz generators \eqref{underlined twistor gen} are invariant under the replacement $\underline{\S}_{\ua} \rightarrow \S_{\ua}\,, \tilde{\underline{\S}}_{\ua}\rightarrow \tilde{\S}_{\ua}\,,$ we can unambiguously write them without an underline.
Explicitly, they take the form 
\bsubeq\label{6d Lorentz gen}
\begin{align}
	(\S_{ab})_{\ah}{}^{\bh}
	&= \left(\begin{array}{cc}
		(\s_{ab})_{\a}{}^{\b} & 0
		\\
		0 & (\tilde{\s}_{ab})^{\ad}{}_{\bd}
	\end{array}\right)\,, & (\S_{a4})_{\ah}{}^{\bh} = \left(\begin{array}{cc}
		0 & -\frac{\ri}{2}(\s_{a})_{\a\bd} \\
		\frac{\ri}{2}(\tilde{\s}_{a})^{\ad\b} & 0
	\end{array}\right)\,, \label{underlined 5d generators}
	\\
	(\S_{a5})_{\ah}{}^{\bh} &= \left(\begin{array}{cc}
		0 & -\frac{\ri}{2}(\s_{a})_{\a\bd} \\
		-\frac{\ri}{2}(\tilde{\s}_{a})^{\ad\b} & 0
	\end{array}\right)\,, 
	& (\S_{45})_{\ah}{}^{\bh} = 
	\left(\begin{array}{cc}
		-\frac{1}{2}\d_{\a}{}^{\b} & 0 \\
		0 & \frac{1}{2}\d^{\ad}{}_{\bd} 
	\end{array}\right) \,.
\end{align}
\esubeq
Here we note that we can express a general Lie algebra element as 
\bsubeq
\begin{align}
	\o = \frac{1}{2}\o^{\ua\ub}(\S_{\ua\ub})\,, \qquad \o^{\ua\ub} &= (\o^{\ua\ub})^{*}\,,
	\\
	\g_{0}\o^{\dag} + \o\g_{0} &= 0\,.
\end{align}
\esubeq
We then have the following
\begin{align} \label{su(2,2) generic algebra element}
	\frac{1}{2}\o^{\ua\ub}(\S_{\ua\ub})_{\ah}{}^{\bh} = 
	\left(\begin{array}{c|c}
		\frac{1}{2}\o^{ab}(\s_{ab})_{\a}{}^{\b} - \frac{1}{2}\o^{45}\d_{\a}{}^{\b} & - \frac{\ri}{2}(\o^{a4} + \o^{a5})(\s_{a})_{\a\bd}\\
		\hline 
		\frac{\ri}{2}(\o^{a4} - \o^{a5})(\tilde{\s}_{a})^{\ad\b} & \frac{1}{2}\o^{ab}(\tilde{\s}_{ab})^{\ad}{}_{\bd} + \frac{1}{2}\o^{45}\d^{\ad}{}_{\bd}
	\end{array}\right)\,.
\end{align}

The matrices \eqref{new sigma def} satisfy a number of additional useful properties.
Their Hermitian conjugation properties can be read off using those of $\underline{\S}_{\ua}$ and $\underline{\tilde{\S}}_{\ua}$ as well as the properties of $\mathcal{C}$ and $\g_{5}$:
\begin{align} \label{new hermit}
	(\S_{\ua})^{\dag} 
	&= \g_{0} \tilde{\S}_{\ua} \g_{0}\,. 
\end{align}
We have used $\mathcal{C}^{\dag} = \mathcal{C}^{-1}$, $\g_{0}^{\T} = \g_{0}$ and \eqref{4d charge conjugation}. 
They also satisfy the following completeness relations
\bsubeq \label{completeness relations}
\begin{align}
	\frac{1}{2}\ve_{\ah\bh\gh\dhat}(\tilde{\S}^{\ua})^{\gh\dhat} &= 
	(\S^{\ua})_{\ah\bh}\,,
	\\
	\frac{1}{2}\ve^{\ah\bh\gh\dhat}(\S^{\ua})_{\gh\dhat} &= 
	(\tilde{\S}^{\ua})^{\ah\bh}  \,,
	\\
	(\S^{\ua})_{\ah\bh}(\S_{\ua})_{\gh\dhat} &= 2\ve_{\ah\bh\gh\dhat}\,.
\end{align}
\esubeq
We can now establish the one-to-one correspondence between complex vectors $V^{\ua}$ in 
${\mathbb R}^{4,2}$
and antisymmetric bi-twistors $V_{\ah\bh} = -V_{\bh\ah}\,,$
\bsubeq \label{o-to-o}
\begin{align}
	V_{\ah\bh} &= V^{\ua}(\S_{\ua})_{\ah\bh}\,, \qquad V^{\ua} = \frac{1}{4}(\tilde{\S}^{\ua})^{\ah\bh}V_{\ah\bh}\,,
	\\
	V^{\ah\bh} &= V^{\ua}(\tilde{\S}_{\ua})^{\ah\bh}\,, \qquad V^{\ua} = \frac{1}{4}(\S^{\ua})_{\ah\bh}V^{\ah\bh}\,. 
\end{align}
\esubeq
In the above, we have 
\begin{align}
	V^{\ah\bh} = \frac{1}{2}\ve^{\ah\bh\gh\dhat}V_{\gh\dhat}\,,
\end{align}
consistent with \eqref{completeness relations}. 
We will now introduce the condensed notation 
\begin{align}
	V := (V_{\ah\bh})\,, \qquad \tilde{V} := (V^{\ah\bh})\,. 
\end{align}
We wish to investigate how the bi-twistor associated with $V^{\ua}$ is related to that associated with $\bar{V}^{\ua}$, the complex conjugate of the six-vector $V^{\ua}\,.$ 
Taking the Hermitian conjugate of $V$, using \eqref{new hermit}, we find
\begin{align} \label{bt real step 1}
	V = V^{\ua}(\S_{\ua}) \rightarrow V^{\dag}
	&= \bar{V}^{\ua}\g_{0}(\tilde{\S}_{\ua})\g_{0}
	\\
	\notag 
	&= \g_{0}\tilde{\bar{V}}\g_{0}\,.
\end{align}
Inserting $V^{\ua} = \bar{V}^{\ua}$ into the above, we find
\bsubeq \label{reality cond for six vector}
\begin{align} 
	V^{\dag} = \g_{0}\tilde{V}\g_{0} \implies \tilde{V} = \g_{0}V^{\dag}\g_{0}\,. 
\end{align}
Now, identifying $\g_{0}$ with $\O$ in \eqref{su(2,2) master} and expressing the reality condition in index notation, we have
\begin{align} \label{index reality}
	\O^{\ah\gh}(V^{\dag})_{\gh\dhat}\O^{\dhat\bh} = \frac{1}{2}\ve^{\ah\bh\gh\dhat}V_{\gh\dhat}\,. 
\end{align}
\esubeq

\subsection{Spinors in $4+1$ dimensions} \label{5d conventions}

Consider the space $\mathbb{R}^{4,1}$ equipped with metric
\begin{align}
	\eta_{\ha\hb} = \text{diag}(-1\,,+1\,,+1\,,+1\,,+1)\,, \qquad \qquad \ha = 0,1,2,3,4\,. 
\end{align}
The corresponding gamma matrices satisfy
\begin{align} \label{5d gm}
	\{\g_{\ha}\,, \g_{\hb}\} = -2\eta_{\ha\hb} \id_{4}\,, \qquad (\g_{\ha})^{\dag} = \g_{0}\g_{\ha}\g_{0}\,, 
\end{align}
and can be chosen as 
\begin{align} \label{5d gamma}
	(\g_{a})_{\ah}{}^{\bh} = \left(\begin{array}{cc}
		0 & (\s_{a})_{\a\bd} \\
		(\tilde{\s}_{a})^{\ad\b} & 0
	\end{array}\right) \,, 
\qquad 
(\g_{4})_{\ah}{}^{\bh} = \left(\begin{array}{cc}
	-\ri \d_{\a}{}^{\b} & 0
	\\
	0 & \ri \d^{\ad}{}_{\bd} 
\end{array} \right) \,.
\end{align}
They are related to the Lorentz spinor generators in ${\mathbb R}^{4,2}$, \eqref{6d Lorentz gen}, as 
\begin{align}
	(\S_{\ha5})_{\ah}{}^{\bh} &= -\frac{\ri}{2}(\g_{\ha})_{\ah}{}^{\bh}\,.
\end{align}
To minimise confusion between the $4+1$ and $4+2$ cases, we will denote the charge conjugation matrix ${\mathbb R}^{4,1}$ as $C$.
The charge conjugation matrix is defined as 
\begin{align} \label{5d cc}
	C\g_{\ha}C^{-1} = (\g_{\ha})^{\T}\,, \qquad C = (\ve^{\ah\bh}) = \left(\begin{array}{cc}
		\ve^{\a\b} & 0 \\
		0 & -\ve_{\ad\bd}
	\end{array}\right)\,. 
\end{align}
The antisymmetric matrices $\ve^{\ah\bh}$ and its inverse, $\ve_{\ah\bh}$, are used to raise and lower the four-component spinors indices in ${\mathbb R}^{4,1}$.

Now we can define the Lorentz generators for the spinor representation,
\begin{align}
	\S_{\ha\hb} := -\frac{1}{4}[\g_{\ha}\,, \g_{\hb}]\,,
\end{align}
and choose them in the form
\begin{align}
	(\S_{ab})_{\ah}{}^{\bh} = \left(\begin{array}{cc}
		(\s_{ab})_{\a}{}^{\b} & 0 \\
		0 & (\tilde{\s}_{ab})^{\ad}{}_{\bd}
	\end{array}\right)\,, \qquad (\S_{a4})_{\ah}{}^{\bh} = \left(\begin{array}{cc}
	0 & -\frac{\ri}{2}(\s_{a})_{\a\bd} \\
	\frac{\ri}{2}(\tilde{\s}_{a})^{\ad\b} & 0
\end{array}\right)\,. 
\end{align}
We can see that these generators coincide with the $4+1$ subset of those given in \eqref{underlined 5d generators}.
We therefore find an explicit form for a general element of the algebra by `switching off' the $\o^{\ha5}$ components of \eqref{su(2,2) generic algebra element}:
\begin{align} \label{so(4,1) generic algebra element}
	\frac{1}{2}\o^{\ha\hb}(\S_{\ha\hb})_{\ah}{}^{\bh} = 
	\left(\begin{array}{c|c}
		\frac{1}{2}\o^{ab}(\s_{ab})_{\a}{}^{\b} & - \frac{\ri}{2}(\o^{a4})(\s_{a})_{\a\bd}\\
		\hline 
		\frac{\ri}{2}(\o^{a4})(\tilde{\s}_{a})^{\ad\b} & \frac{1}{2}\o^{ab}(\tilde{\s}_{ab})^{\ad}{}_{\bd}
	\end{array}\right)\,.
\end{align}

Given \eqref{5d gm} and \eqref{5d cc}, we have the following properties
\begin{align}
	(\S_{\ha\hb})^{\dag} = -\g_{0}(\S_{\ha\hb})\g_{0}\,, \qquad (\S_{\ha\hb})^{\T} = - C(\S_{\ha\hb})C^{-1}\,.
\end{align}
Writing an element of the Lie algebra $\mathfrak{so}(4,1)$ as \eqref{so(4,1) generic algebra element}:
\begin{align}
	\o := \frac{1}{2}\o^{\ha\hb}(\S_{\ha\hb})\,, \qquad \o^{\ha\hb} = (\o^{\ha\hb})^{*}\,,
\end{align}
we find the following master equations
\bsubeq
\begin{align} \label{so(4,1) master algebra}
	C\o + \o^{\T}C &= 0\,, 
	\\
	\g_{0}\o + \o^{\dag}\g_{0} &= 0\,.
\end{align}
\esubeq
These correspond to the group equations 
\bsubeq \label{so(4,1) master group}
\begin{align}
	g^{\T}Cg &= C \,,
	\\
	g^{\dag}\g_{0}g &= \g_{0}\,.
\end{align}
\esubeq
This group is known as $\sUSp(2,2)$. 

A Dirac spinor in ${\mathbb R}^{4,1}$, $\Psi = (\Psi_{\ah})\,,$ and its Dirac conjugate, $\bar{\Psi} = (\bar{\Psi}^{\ah}) = \Psi^{\dag}\g_{0}\,,$ look like
\begin{align}
	\Psi_{\ah} = \left(\begin{array}{c}
		\psi_{\a} \\
		\bar{\f}^{\ad}
	\end{array}\right) \,, \qquad \bar{\Psi}^{\ah} = \left(\begin{array}{cc}
	\f^{\a}\,, & \bar{\psi}_{\ad}
\end{array}\right)\,. 
\end{align}
We can combine $\bar{\Psi}^{\ah} = \left(\f^{\a}\,, \bar{\psi}_{\ad}\right)$ and $\Psi^{\ah} = \ve^{\ah\bh}\Psi_{\bh} = \left(\psi^{\a}\,, - \bar{\f}_{\ad}\right)$ into an $\sSU(2)$ doublet:
\begin{align}
	\Psi_{\ui}^{\ah} = \left(\Psi^{\a}_{\ui}\,, - \bar{\Psi}_{\ad \ui}\right) \,,
	\qquad \left(\Psi_{\ui}^{\a}\right)^{*} = \bar{\Psi}^{\ad \ui}\,, \qquad \ui = 1\,,2\,,
\end{align}
such that $\Psi_{\underline{1}}^{\a} = \f^{\a}$ and $\Psi_{\underline{2}}^{\a} = \psi^{\a}\,.$ The $\sSU(2)$ indices are raised and lowered in the usual way: $\Psi^{\ah \ui} = \ve^{\ui\uj}\Psi^{\ah}_{\uj}\,.$ The spinor $\Psi^{\ui} = (\Psi_{\ah}^{\ui})$ satisfies the pseudo-Majorana condition 
\begin{align} \label{pseudo majorana}
	\bar{\Psi}_{\ui}^{\T} = C\Psi_{\ui}\,. 
\end{align}

\begin{footnotesize}

\end{footnotesize}

\end{document}